%% file: sss04.tex
\documentclass[aps,prd,preprint,tightenlines,superscriptaddress,showpacs,byrevtex]{revtex4}
\usepackage{graphicx}
\usepackage{dcolumn}
\usepackage{bm}
\usepackage{pstricks}
\usepackage{pst-node}

\begin{document}

\preprint{\vbox{
                 \hbox{BELLE-CONF-0435}
                 \hbox{ICHEP04 6-0681}
}}

\input{defs.tex}
\input{sss04_defs.tex}

\title{\quad\\[0.5cm] \boldmath New Measurements of Time-Dependent {\boldmath $CP$}-Violating 
       \\Asymmetries in $b \to s$ Transitions at Belle}

\date{\today}

\input{ichep04authors.tex}

\begin{abstract}
  We present new measurements of $CP$-violation parameters
  in 
  $\bz\to$
  $\phi\kz$, 
  $\kp\km\ks$,
  $\fzero\ks$,
  $\eta'\ks$, 
  $\omega\ks$, 
  $\ks\piz$, 
  and
  $\kstarz\gamma~(\kstarz\to\ks\piz)$
  decays
  based on a sample of $275\times 10^6$ $B\bbar$ pairs
  collected at the $\ufs$ resonance with
  the Belle detector at the KEKB energy-asymmetric $e^+e^-$ collider.
  One neutral $B$ meson is fully reconstructed in
  one of the specified decay channels,
  and the flavor of the accompanying $B$ meson is identified from
  its decay products.
  $CP$-violation parameters for each of the decay 
  modes are obtained from the asymmetries in the distributions of
  the proper-time intervals between the two $B$ decays.
  All results are preliminary.
\end{abstract}


\maketitle

\section{Introduction}
\label{sec:introduction}

The phenomena of $CP$ violation in the flavor-changing $b \to s$ transition
are sensitive to physics at a very high-energy scale~\cite{Akeroyd:2004mj}.
Theoretical studies indicate that
large deviations from standard model (SM) expectations
are allowed for time-dependent $CP$ asymmetries in
$\bz$ meson decays~\cite{bib:lucy}.
Experimental investigations have recently been launched
at the two $B$ factories, each of which has produced more than
$10^8$ $B\bbar$ pairs.
Belle's previous measurement of the $\bz\to \phi\ks$ decay~\cite{bib:CC},
which is dominated by the $\btosss$ transition,
yielded a value that differs from the SM expectation 
by 3.5 standard deviations~\cite{Abe:2003yt}.
Measurements with a larger data sample are required to 
elucidate this difference. It is also essential to examine
additional modes that
may be sensitive to the same $b \to s$ penguin amplitude.
In this spirit, experimental results using the decay modes
$\bz\to\phi\kl$, 
$\kp\km\ks$, 
$\fzero\ks$,
$\eta'\ks$, 
and
$\ks\piz$ have already been reported~\cite{Abe:2003yt,bib:BaBar_sss}.

In the SM, $CP$ violation arises from an
irreducible phase, the Kobayashi-Maskawa (KM) phase~\cite{Kobayashi:1973fv},
in the weak-interaction quark-mixing matrix. In
particular, the SM predicts $CP$ asymmetries in the time-dependent
rates for $\bz$ and
$\bzb$ decays to a common $CP$ eigenstate $\fCP$~\cite{bib:sanda}. 
In the decay chain $\Upsilon(4S)\to \bz\bzb \to f_{CP}f_{\rm tag}$,
where one of the $B$ mesons decays at time $t_{CP}$ to a 
final state $f_{CP}$ 
and the other decays at time $t_{\rm tag}$ to a final state  
$f_{\rm tag}$ that distinguishes between $B^0$ and $\bzb$, 
the decay rate has a time dependence
given by
\begin{equation}
\label{eq:psig}
{\cal P}(\Delta{t}) = 
\frac{e^{-|\Delta{t}|/{\taubz}}}{4{\taubz}}
\biggl\{1 + \fq\cdot 
\Bigl[ \cals\sin(\dmd\Delta{t})
   + \cala\cos(\dmd\Delta{t})
\Bigr]
\biggr\}.
\end{equation}
Here $\cals$ and $\cala$ are $CP$-violation parameters,
$\taubz$ is the $B^0$ lifetime, $\dmd$ is the mass difference 
between the two $B^0$ mass
eigenstates, $\Delta{t}$ = $t_{CP}$ $-$ $t_{\rm tag}$, and
the $b$-flavor charge $\fq$ = +1 ($-1$) when the tagging $B$ meson
is a $B^0$ ($\bzb$).
To a good approximation,
the SM predicts $\cals = -\xi_f\sin 2\phi_1$, where $\xi_f = +1 (-1)$ 
corresponds to  $CP$-even (-odd) final states, and $\cala =0$
for both $b \to c\overline{c}s$ and 
$b \to s\overline{s}s$ transitions.
Recent measurements of time-dependent $CP$ asymmetries in
$\bz \to J/\psi \ks$ and related decay 
modes, which are governed by the $b \to c\overline{c}s$ transition,
by Belle~\cite{bib:CP1_Belle,bib:BELLE-CONF-0436}
and BaBar~\cite{bib:CP1_BaBar}
already determine $\sinbb$ rather precisely;
the present world average value is 
$\sinbb = \sinbbWAResult$~\cite{bib:HFAG}.
This serves as a firm reference point for the SM.

Belle's previous measurements
for $\bz \to \phi\ks$, $K^+K^-\ks$ and $\eta'\ks$
were based on a 140 fb$^{-1}$ data sample (DS-I)
containing $152\times 10^6$ $B\bbar$ pairs.
While $\phi\ks$ and $\eta'\ks$ final states are
$CP$ eigenstates with $\xi_f=-1$, 
the $K^+K^-\ks$ final state is in general
a mixture of both $\xi_f=+1$ and $-1$.
Excluding $K^+K^-$ pairs that are consistent 
with a $\phi \to K^+K^-$ decay from the $\bz \to K^+K^-\ks$ sample,
we find that the $K^+K^-\ks$ state is primarily $\xi_f=+1$;
a measurement of the $\xi_f=+1$ fraction with DS-I gives
$1.03 \pm 0.15\mbox{(stat)}\pm0.05\mbox{(syst)}$~\cite{Abe:2003yt}.
In the following determination of $\cals$ and $\cala$,
we fix $\xi_f=+1$ for this mode.

In this report, we describe improved measurements 
incorporating an additional 113 fb$^{-1}$
data sample that contains $123\times 10^6$ $B\bbar$
pairs (DS-II) for a total of $275\times 10^6$ $B\bbar$ pairs.
We include additional $\phi\ks$ and $\eta'\ks$
subdecay modes that were not used in the previous analysis.
We also describe new measurements of $CP$ asymmetries for
the following $CP$-eigenstate $\bz$ decay modes:
$\bz\to \phi\kl$ and $\fzero\ks$ for $\xi_f = +1$;
$\bz\to\omega\ks$ 
and
$\ks\piz$
for $\xi_f = -1$.
The decays $\bz\to\phi\ks$ and $\phi\kl$ are combined
in this analysis by redefining $\cals$ as $-\xi_f\cals$ to take
the opposite $CP$ parities into account, 
and are collectively called ``$\bz\to\phi\kz$''.
The $CP$ asymmetries for the decay $\bz\to\omega\ks$
are measured for the first time.

Finally, we also measure time-dependent $CP$ violation in
the decay $\bz\to\kstarz\gamma~(\kstarz\to\ks\piz)$~\cite{bib:Kstargamma_tcpv_BaBar}, 
which is not a $CP$ eigenstate but is sensitive to
physics beyond the SM~\cite{Atwood:1997zr}.
Within the SM, the photon emitted from a $\bz$ ($\bzb$)
meson is dominantly right-handed (left-handed).
Therefore the polarization of the photon carries
information on the original $b$-flavor; the decay
is thus almost flavor-specific. The SM predicts
a small asymmetry $\cals \sim -2(m_s/m_b)\sinbb$, where $m_b$ ($m_s$) is
the $b$-quark ($s$-quark) mass~\cite{Atwood:1997zr}.
Any significant deviation from this expectation
would be a manifestation of physics beyond the SM.

At the KEKB energy-asymmetric 
$e^+e^-$ (3.5 on 8.0~GeV) collider~\cite{bib:KEKB},
the $\Upsilon(4S)$ is produced
with a Lorentz boost of $\beta\gamma=0.425$ nearly along
the electron beamline ($z$).
Since the $B^0$ and $\bzb$ mesons are approximately at 
rest in the $\Upsilon(4S)$ center-of-mass system (cms),
$\Delta t$ can be determined from the displacement in $z$ 
between the $f_{CP}$ and $f_{\rm tag}$ decay vertices:
$\Delta t \simeq (z_{CP} - z_{\rm tag})/(\beta\gamma c)
 \equiv \Delta z/(\beta\gamma c)$.

The Belle detector is a large-solid-angle magnetic
spectrometer that
consists of a silicon vertex detector (SVD),
a 50-layer central drift chamber (CDC), an array of
aerogel threshold Cherenkov counters (ACC),
a barrel-like arrangement of time-of-flight
scintillation counters (TOF), and an electromagnetic calorimeter
comprised of CsI(Tl) crystals (ECL) located inside
a superconducting solenoid coil that provides a 1.5~T
magnetic field.  An iron flux-return located outside of
the coil is instrumented to detect $K_L^0$ mesons and to identify
muons (KLM).  The detector
is described in detail elsewhere~\cite{Belle}.
Two inner detector configurations were used. A 2.0 cm radius beampipe
and a 3-layer silicon vertex detector (SVD-I) were used for DS-I,
while a 1.5 cm radius beampipe, a 4-layer
silicon detector (SVD-II) and a small-cell inner drift chamber were used for 
DS-II~\cite{Ushiroda}.

\section{Event Selection, Flavor Tagging and Vertex Reconstruction}
\subsection{Overview}
We reconstruct the following $\bz$ decay modes to
measure $CP$ asymmetries:
$\bz\to\phi\ks$, 
$\phi\kl$, 
$\kp\km\ks$, 
$\fzero\ks$,
$\eta'\ks$,
$\omega\ks$, 
$\ks\piz$, 
and
$\kstarz\gamma~(\kstarz\to\ks\piz)$.
We exclude $K^+K^-$ pairs that are consistent with a $\phi \to K^+K^-$ decay 
from the $\bz \to K^+K^-\ks$ sample.
The intermediate meson states are reconstructed from the following decays:
$\piz\to\gamma\gamma$,
$\ks \to \pip\pim$ (also $\piz\piz$ for the $\phi\ks$ decay), 
$\eta\to\gamma\gamma$ or $\pip\pim\piz$, 
$\rhoz\to\pip\pim$,
$\omega\to\pip\pim\piz$,
$\kstarz\to\ks\piz$,
$\eta'\to\rhoz\gamma$ or $\eta\pip\pim$,
$\fzero\to\pip\pim$,
and $\phi\to K^+K^-$.

Among the decay chains listed above,
$\bz\to\phi\ks~(\ks\to\pip\pim)$,
$\bz\to\kp\km\ks$,
$\bz\to\eta'\ks~(\eta'\to\rhoz\gamma)$, and
$\bz\to\eta'\ks~(\eta'\to\eta\pip\pim,~\eta\to\gamma\gamma)$
decays were used in the previous analysis~\cite{Abe:2003yt}.
The selection criteria for these decays remain the same.
For the other $\bz$ decay modes that are now included,
identification of photons, neutral and charged kaons, 
and neutral and charged pions is based on
the procedure used previously.
However, the selection criteria
for each $\bz$ decay mode were optimized individually and
are thus different from one another.

\subsection{\boldmath $\bz\to\phi\ks$ and $\kp\km\ks$}
We use well-reconstructed charged tracks with a sufficient
number of associated hits in the CDC.
Charged kaons and pions except for those from $\ks\to\pip\pim$ decays
are required to originate from the interaction point (IP).
We distinguish charged kaons from pions based on
a kaon (pion) likelihood $\mathcal{L}_{K(\pi)}$
derived from the TOF, ACC and $dE/dx$ measurements in the CDC.

Pairs of oppositely charged tracks that have an invariant mass
within 0.030 GeV/$c^2$
of the nominal $\ks$ mass are used to reconstruct $\ks\to\pip\pim$ decays.
The $\pip\pim$ vertex is required to be displaced from
the IP by a minimum transverse distance of 0.22~cm for
high momentum ($>1.5$ GeV/$c$) candidates and 0.08~cm for
those with momentum less than 1.5~GeV/$c$.
The direction of the pion pair momentum must also agree with
the direction defined by the IP and the vertex displacement
within 0.03 rad for high-momentum candidates, and within 0.1
rad for the remaining candidates.

Photons are identified as isolated ECL clusters
that are not matched to any charged track.
To select $\ks\to\piz\piz$ decays,
we reconstruct $\piz$ candidates from pairs of photons
with $E_\gamma > 0.05$ GeV, where
$E_\gamma$ is the photon energy measured with the ECL.
The reconstructed $\piz$ candidate is required to
have an invariant mass between  0.08 and 0.15 GeV$/c^2$
and a momentum above 0.1 GeV/$c$.
The large mass range is used to achieve a high reconstruction efficiency.
Candidate $\ks\to\piz\piz$ decays are required to have
invariant masses between 0.47 GeV/$c^2$ and 0.52 GeV/$c^2$,
where we perform a fit with constraints on
the $\ks$ vertex and the $\piz$ masses
to improve the $\piz\piz$ invariant mass resolution.
We also require that the distance between the IP and the
reconstructed $\ks$ decay vertex be larger than $-10$~cm,
where the positive direction is defined by the $\ks$ momentum.

Candidate $\phi \to \kp\km$ decays are required to
have a $\kp\km$ invariant mass that is within 0.01 GeV/$c^2$ 
of the nominal $\phi$ meson mass.
Since the $\phi$ meson selection is effective in reducing background events,
we impose only minimal kaon-identification requirements;
$\rkpi \equiv \mathcal{L}_K /(\mathcal{L}_K + \mathcal{L}_\pi) > 0.1$
is required, 
where the kaon likelihood ratio $\mathcal{R}_{K/\pi}$ has values 
between 0 (likely to be a pion) and 1 (likely to be a kaon).
We use a more stringent kaon-identification requirement,
$\rkpi > 0.6$,
to select non-resonant $\kp\km$ candidates 
for the decay $\bz \to \kp\km\ks$.
We reject $K^+K^-$ pairs that are consistent with 
$D^0 \to K^+K^-$, $\chi_{c0} \to K^+K^-$, or $J/\psi \to K^+K^-$ decays.
We also remove $D^+ \to \ks K^+$ candidates.

For reconstructed $B\to\fCP$ candidates, we identify $B$ meson decays using the
energy difference $\dE\equiv E_B^{\rm cms}-E_{\rm beam}^{\rm cms}$ and
the beam-energy constrained mass $\mb\equiv\sqrt{(E_{\rm beam}^{\rm cms})^2-
(p_B^{\rm cms})^2}$, where $E_{\rm beam}^{\rm cms}$ is
the beam energy in the cms, and
$E_B^{\rm cms}$ and $p_B^{\rm cms}$ are the cms energy and momentum of the 
reconstructed $B$ candidate, respectively.
The $B$ meson signal region is defined as 
$|\dE|<0.06$ GeV for $\bz \to \phi \ks~(\ks\to\pip\pim)$,
$-0.15~{\rm GeV} < \dE < 0.1$ GeV for $\bz \to \phi \ks~(\ks\to\piz\piz)$,
$|\dE|<0.04$ GeV for $\bz \to K^+K^-\ks$,
and $5.27~{\rm GeV}/c^2 <\mb<5.29~{\rm GeV}/c^2$ for all decays.

The dominant background to the $\bz\to\phi\ks$ decay comes from
$e^+e^- \rightarrow 
u\overline{u},~d\overline{d},~s\overline{s}$, or $c\overline{c}$
continuum events. Since these tend to be jet-like, while
the signal events tend to be spherical,
we use a set of variables that characterize the event topology
to distinguish between the two.
We combine 
$\sperp$, $\theta_T$ and modified Fox-Wolfram moments~\cite{Abe:2001nq}
into a Fisher discriminant $\calf$, where
$\sperp$ is
the scalar sum of the transverse momenta of all
particles outside a $45^\circ$ cone around the candidate $\phi$
meson direction divided by the scalar sum of their total momenta,
and
$\theta_T$ is
the angle between the thrust axis of the $B$ candidate
and that of the other particles in the cms.
We also use the angle of the reconstructed $\bz$
candidate with respect to the beam direction in the cms
($\theta_B$), 
and
the helicity angle $\theta_H$ defined as the
angle between the $\bz$ meson momentum and the daughter
$\kp$ momentum in the $\phi$ meson rest frame.
We combine
$\calf$, $\cos\theta_B$ and $\cos\theta_H$
into a signal [background]
likelihood variable, which is defined as 
${\cal L}_{\rm sig[bkg]} \equiv
{\cal L}_{\rm sig[bkg]}(\calf)\times
{\cal L}_{\rm sig[bkg]}(\cos\theta_B)\times
{\cal L}_{\rm sig[bkg]}(\cos\theta_H)$.
We impose requirements on the likelihood ratio 
$\rsigbkg \equiv \lsig/(\lsig+\lbkg)$ to
maximize the figure-of-merit (FoM) defined as
$\nsigmc/\sqrt{\nsigmc+\nbkg}$, where $\nsigmc$ ($\nbkg$) 
represents the expected
number of signal (background) events in the signal region.
We estimate $\nsigmc$ using Monte Carlo (MC) events, while
$\nbkg$ is determined from events outside the signal region.
The requirement for $\rsigbkg$
depends both on the decay mode
and on the flavor-tagging quality, $r$, which is
described in Sec.~\ref{sec:flavor tagging}.
The threshold values range from 0.1 (used for $r>0.875$)
to 0.4 (used for $r<0.25$) for the decay
$\bz\to\phi\ks~(\ks\to\pip\pim)$,
and from 0.25 to 0.65 for the decay $\bz\to\kp\km\ks$.
We impose a more stringent requirement, $\rsigbkg > 0.75$,
for all $r$ values
in the decay $\bz\to\phi\ks~(\ks\to\piz\piz)$.

We use events outside the signal region 
as well as a large MC sample to study the background components.
The dominant background is from continuum.
The contributions from $B\overline{B}$ events are small.
The contamination of $\bz\to\kp\km\ks$ events in the $\bz\to\phi\ks$ sample 
is $7.1\pm1.6$\% ($6.2\pm2.0$\%) for DS-I (DS-II).
Backgrounds from the decay $\bz\to\fzero\ks~(\fzero\to\kp\km)$, 
which has a
$CP$ eigenvalue opposite to $\phi\ks$, are found to be 
$0.4^{+1.9}_{-0.4}$\% ($0.0^{+2.0}_{-0.0}$\%) for DS-I (DS-II).
The influence of these backgrounds
is treated as a source of systematic uncertainty.

Figure~\ref{fig:mb}(a) and (c) show
the $\mb$ distribution for the reconstructed $\bz\to\phi\ks$ 
and $\kp\km\ks$ candidates
within the $\dE$ signal regions
after flavor tagging and vertex reconstruction.
The signal yield is determined
from an unbinned two-dimensional maximum-likelihood fit
to the $\dE$-$\mb$ distribution in the fit region
defined as
$ 5.2~{\rm GeV/}c^2 < \mb < 5.3~{\rm GeV/}c^2$ 
for all modes, and
$-0.12~{\rm GeV} < \dE < 0.25~{\rm GeV}$ for the
$\bz\to\phi\ks~(\ks\to\pip\pim)$ or $\kp\km\ks$ decay and
$-0.15~{\rm GeV} < \dE < 0.25~{\rm GeV}$ for the
$\bz\to\phi\ks~(\ks\to\piz\piz)$ decay.
The $\phi\ks~(\ks\to\pip\pim)$ signal distribution 
is modeled with a Gaussian function (a sum of two Gaussian functions)
for $\mb$ ($\dE$).
The $\phi\ks~(\ks\to\piz\piz)$ signal distribution 
is modeled with a smoothed histogram obtained from MC events.
For the continuum background,
we use the ARGUS parameterization~\cite{bib:ARGUS} 
for $\mb$
and a linear function for $\dE$.
The fits yield $\Nsigphiks$ $\bz\to\phi\ks$ events and
$\Nsigkpkmks$ $\bz\to\kp\km\ks$ events in the signal region,
where the errors are statistical only.

\subsection{\boldmath $\bz\to\phi\kl$}
\label{sec:bztophikl}
Candidate $\phi \to \kp\km$ decays
are selected with the same criteria as described above. 
We select $\kl$ candidates based on KLM and ECL
information. There are two classes of $\kl$ candidates
that we refer to as KLM and ECL candidates.
The requirements for the KLM candidates are the same
as those used in the $\bz\to\jpsi\kl$ selection
for the $\sinbb$ measurement~\cite{bib:CP1_Belle}.
ECL candidates are selected from ECL clusters using
a $\kl$ likelihood ratio~\cite{bib:CP1_Belle}, which is
calculated from
the following information: the distance between the
ECL cluster and the closest extrapolated charged track position;
the ECL cluster energy; $E_9/E_{25}$, the ratio of energies summed in
$3\times 3$ and $5\times 5$ arrays of CsI(Tl) crystals surrounding
the crystal at the center of the shower; the ECL shower
width and the invariant mass of the shower. 
The likelihood ratio is required to be greater than 0.8.
For both KLM and ECL candidates, we also require
that the cosine of the angle between the $\kl$ direction
and the direction of the missing momentum of the event
in the laboratory frame be greater than 0.6.

Since the energy of the $\kl$ is not measured,
$\mb$ and $\dE$ cannot be calculated
in the same way that is used for the other final states.
Using the four-momentum of a reconstructed
$\phi$ candidate and the $\kl$ flight direction,
we calculate the momentum of the $\kl$ candidate
requiring $\dE=0$. We then calculate
$\pbstar$, the momentum of the $B$ candidate in the
cms, and define the
$B$ meson signal region as 
$0.2~{\rm GeV/}c < \pbstar < 0.5~{\rm GeV/}c$.
We impose the requirement $\rsigbkg > 0.98$ to reduce the continuum background.
Here $\rsigbkg$ is based on the discriminating variables
used for the $\bz\to\phi\ks$ decay and the number of
tracks originating from the IP with a momentum above 0.1 GeV/$c$.
The $\rsigbkg$ requirement is chosen to optimize the FoM, which is
calculated taking the background from
both continuum and generic $B$ decays into account.
The $\kl$ detection efficiency difference between data and MC 
is studied using the decay $\bz\to\jpsi\kl$, and corrections
are applied to the $\bz\to\phi\kl$ MC events to calculate
the FoM.
If there is more than one candidate $\bz\to\phi\kl$ decay
in the signal region,
we take the one with the highest $\rsigbkg$ value.
ECL candidates are not used if there is a candidate
$\bz\to\phi\kl$ decay with a KLM candidate.
We find that about 90\% of signal events are reconstructed
with KLM candidates.

We study the background components
using a large MC sample
as well as data taken with cms energy 60 MeV
below the nominal $\Upsilon(4S)$ mass (off-resonance data).
The dominant background is from continuum.
A MC study with the efficiency correction obtained from
$\bz\to\jpsi\kl$ data yields
$9\pm 5$ background events
from $B$ decays, which include
$\bz\to\phi\kstarz$, $\phi\ks$ and $\bp\to\phi\kstarp$
decays.
The influence of these backgrounds, including
their $CP$ asymmetries,  is treated as a source
of systematic uncertainty.

The $\pbstar$ distribution after flavor tagging 
and vertex reconstruction
is shown in Fig.~\ref{fig:mb}(b).
The signal yield is determined from
an extended unbinned maximum-likelihood fit in the range
$0~{\rm GeV/}c < \pbstar < 1~{\rm GeV/}c$.
The $\bz\to\phi\kl$ signal shape is obtained from MC events.
Background from $B\bbar$ pairs is also modeled with
MC. We fix the ratio between the signal yield
and the $B\bbar$ background based on
known branching fractions and reconstruction
efficiencies; uncertainty in the ratio is treated
as a source of systematic error.
The continuum background distribution is represented
by a smoothed histogram obtained from MC events;
we confirm that the function describes
the off-resonance data well.
The fit yields $\Nsigphikl\pm 10$ $\bz\to\phi\kl$ events,
where the first error is statistical and the second
error is systematic.
The sources of the systematic error include
uncertainties in the efficiency corrections,
in $B\bbar$ background branching fractions and
in the background parameterizations.
The result is in good agreement with
the expected $\bz\to\phi\kl$ signal yield 
($36\pm 9$ events) obtained from MC after applying
the efficiency correction from the
$\bz\to\jpsi\kl$ data.

\subsection{\boldmath $\bz\to\fzero\ks$}
Candidate $\ks\to\pip\pim$ decays are selected
with the criteria that are slightly different from
those used for the $\bz\to\phi\ks$ decay
to obtain the best performance for the $\bz\to\fzero\ks$ decay. 
Pairs of oppositely charged pions 
that have invariant masses
between 0.890 and 1.088 GeV/$c^2$ are used to reconstruct
$\fzero\to\pip\pim$ decays.
Tracks that are identified as kaons ($\rkpi > 0.7$) or
electrons are not used.
We require that both $\ks\pip$ and $\ks\pim$
combinations have invariant masses more than
0.1 GeV/$c^2$ above the nominal charged
$D$ meson mass; this
removes background from $D^\pm\to\ks\pi^\pm$ and 
$K^{*\pm}\to\ks\pi^\pm$ decays.

The $B$ meson signal region is defined as 
$|\dE|<0.06$ GeV and $5.27~{\rm GeV}/c^2 <\mb<5.29~{\rm GeV}/c^2$.
The dominant background is from continuum.
For the continuum suppression, we require
$\rsigbkg > 0.6$ for events with the best-quality flavor tagging
($r > 0.875$), and $\rsigbkg > 0.8$ for other events.
Here the signal likelihood ratio $\rsigbkg$ is obtained from
$\cos\theta_B$ and
$\calf$, which consists of the modified Fox-Wolfram moments
and $\cos\theta_T$.

Figure~\ref{fig:mb}(d) shows
the $\mb$ distribution for the reconstructed $\bz\to\fzero\ks$ candidates
within the $\dE$ signal region
after flavor tagging and vertex reconstruction.
For the signal yield extraction,
we first perform
an unbinned two-dimensional maximum-likelihood fit
to the $\dE$-$\mb$ distribution in the fit region defined as
$ 5.2~{\rm GeV/}c^2 < \mb < 5.3~{\rm GeV/}c^2$ 
and
$-0.3~{\rm GeV} < \dE < 0.4~{\rm GeV}$.
The signal is modeled with a Gaussian function
(a sum of two Gaussian functions) for $\mb$ ($\dE$).
For the continuum background, 
we use the ARGUS parameterization for $\mb$
and a linear function for $\dE$.
The fit yields the number of 
$\bz\to\pip\pim\ks$ events that have
$\pip\pim$ invariant masses within
the $\fzero$ resonance region, which
may include contributions from
$\bz\to\rhoz\ks$ as well as non-resonant three-body
$\bz\to\pip\pim\ks$ decays.
To separate these peaking backgrounds from
the $\bz\to\fzero\ks$ decay,
we perform another fit to the $\pip\pim$ invariant
mass distribution for the events inside the
$\dE$-$\mb$ signal region. We use Breit-Wigner
functions for the $\bz\to\fzero\ks$ signal as well
as for $\bz\to\rho\ks$ and a possible resonance
above the $\fzero$ mass region~\cite{Garmash:2003er}.
Three-body $\bz\to\pip\pim\ks$ decays are modeled  with
a fourth-order polynomial function. The other background
is modeled with a threshold function.
The fit yields $\Nsigfzeroks$ $\bz\to\fzero\ks$ events.
The peaking background contribution in the 
$\dE$-$\mb$ signal region is estimated to be $9\pm 3$ events.

\subsection{\boldmath $\bz\to\eta'\ks$}
Candidate $\ks \to \pip\pim$ decays
are selected with the same criteria as those used for
the $\bz\to\phi\ks$ decay.
Charged pions from the $\eta$, $\rhoz$ or $\eta'$ decay
are selected from tracks originating from the IP.
We reject kaon candidates by requiring $\rkpi < 0.9$. 
Candidate photons from
$\piz\to\gamma\gamma$ decays are
required to have $E_\gamma > 0.05 $~GeV.
The reconstructed $\piz$ candidate is required to satisfy 
$0.118~{\rm GeV}/c^2 < \mgg < 0.15~{\rm GeV}/c^2$
and
$\ppizcms > 0.1~{\rm GeV}/c$, where
$\mgg$ and $\ppizcms$ are the invariant mass and
the momentum in the cms, respectively. 
Candidate photons from
$\eta\to\gamma\gamma~(\eta'\to\rhoz\gamma)$ decays are
required to have $E_\gamma > 0.05~(0.1)$~GeV.
The invariant mass of the photon pair
is required to be between 0.5 and
0.57~GeV/$c^2$ for the $\eta\to\gamma\gamma$ decay.
The $\pip\pim\piz$ invariant mass is required
to be between 0.535 and 0.558~GeV/$c^2$ for the
$\eta\to\pip\pim\piz$ decay.
A kinematic fit with an $\eta$ mass constraint is
performed using the fitted vertex of the $\pi^+\pi^-$ tracks from
the $\eta^\prime$ as the decay point. 
For $\eta^\prime\to\rhoz\gamma$ decays, candidate $\rhoz$ mesons
are reconstructed from pairs of vertex-constrained $\pi^+\pi^-$
tracks with an invariant mass between 0.55 and 0.92~GeV/$c^2$. 
The $\eta^\prime\to\eta\pip\pim$ candidates are required 
to have a reconstructed mass
between 0.94 and 0.97~GeV/$c^2$ (0.95 and 0.966~GeV/$c^2$)
for the $\eta\to\gamma\gamma$ ($\eta\to\pip\pim\piz$) decay.
Candidate $\eta^\prime\to\rhoz\gamma$ decays are required to
have a reconstructed mass from 0.935 to 0.975~GeV/$c^2$.

The $B$ meson signal region is defined as 
$|\dE|<0.06$ GeV for $\bz \to \eta'\ks~(\eta'\to \rhoz\gamma) $,
$-0.1$ GeV $< \dE <0.08$ GeV for 
$\bz \to \eta'\ks~(\eta'\to\eta\pip\pim,~\eta\to\gamma\gamma) $
or 
$-0.08$ GeV $< \dE <0.06$ GeV for 
$\bz \to \eta'\ks~(\eta'\to\eta\pip\pim,~\eta\to\pip\pim\piz) $,
and $5.27~{\rm GeV}/c^2 <\mb<5.29~{\rm GeV}/c^2$ for all decays.
The continuum suppression is based on the
likelihood ratio $\rsigbkg$ obtained from
the same discriminating variables 
used for the $\bz\to\phi\ks$ decay, except that
$\cos\theta_H$ for $\bz$ decays is not used;
we only use the helicity angle for
the decay $\eta'\to\rho\gamma~(\rho\to\pip\pim)$,
which is defined as
the angle between the $\eta'$ meson momentum and the daughter
$\pip$ momentum in the $\rho$ meson rest frame.
The minimum $\rsigbkg$ requirement
depends both on the decay mode and on the
flavor-tagging quality, and
ranges from 0 (i.e. no requirement) to 0.4.

We use events outside the signal region 
as well as a large MC sample to study the background components.
The dominant background is from continuum.
In addition, according to MC simulation, there is  a small ($\sim 3\%$) 
contamination from $B\overline{B}$ background events
in $\bz\to\eta'\ks~(\eta'\to\rhoz\gamma)$. 
The contributions from $B\overline{B}$ events are smaller for other modes.
The influence of these backgrounds
is treated as a source of systematic uncertainty.

Figure~\ref{fig:mb}(e) shows
the $\mb$ distribution for the reconstructed $\bz\to\eta'\ks$ candidates
within the $\dE$ values signal region
after flavor tagging and vertex reconstruction.
The signal yield is determined
from unbinned two-dimensional maximum-likelihood fits
to the $\dE$-$\mb$ distributions in the fit region defined as
$ 5.2~{\rm GeV/}c^2 < \mb < 5.3~{\rm GeV/}c^2$ 
and
$-0.25~{\rm GeV} < \dE < 0.25~{\rm GeV}$.
We perform the fit for each final state separately.
The $\eta'\ks$ signal distribution 
is modeled with a sum of two (three) Gaussian functions
for $\mb$ ($\dE$).
For the continuum background, 
we use the ARGUS parameterization for $\mb$
and a linear function for $\dE$.
For the $\eta'\to\rho\gamma$ mode, we include
the $B\bbar$ background shape obtained from MC
in the fits.
The fits yield a total of $\Nsigetapks$ $\bz\to\eta'\ks$ events
in the signal region,
where the error is statistical only.

\subsection{\boldmath $\bz\to\omega\ks$}
Candidate $\ks \to \pip\pim$ decays
are selected with criteria that
are identical to those used for the $\bz\to\phi\ks$ decay.
Pions for the $\omega\to\pip\pim\piz$ decay
are selected with the same criteria used
for the $\eta\to\pip\pim\piz$ decay, except that
we require $\ppizcms > 0.35~{\rm GeV}/c$.
The $\pip\pim\piz$ invariant mass is required to be
within 0.03 GeV/$c^2$ of the nominal $\omega$ mass.
The $B$ meson signal region is defined as 
$|\dE|<0.06$ GeV and $5.27~{\rm GeV}/c^2 <\mb<5.29~{\rm GeV}/c^2$.
The dominant background is from continuum.
The continuum suppression is based on the
likelihood ratio $\rsigbkg$ obtained from
the same discriminating variables 
used for the $\bz\to\phi\ks$ decay;
the helicity angle $\theta_H$ is defined as the
angle between the $\bz$ meson momentum and the cross product
of the $\pip$ and $\pim$ momenta
in the $\omega$ meson rest frame.
The minimum $\rsigbkg$ requirement
depends both on the decay mode and on the
flavor-tagging quality, and
ranges from 0.3 (used for $r>0.875$)
to 0.9 (used for $r<0.25$).
The contribution from $B\overline{B}$ events is negligibly small.

Figure~\ref{fig:mb}(f) shows
the $\mb$ distribution for the reconstructed $\bz\to\omega\ks$ candidates
within the $\dE$ signal region
after flavor tagging and vertex reconstruction.
The signal yield is determined
from an unbinned two-dimensional maximum-likelihood fit
to the $\dE$-$\mb$ distribution in the fit region defined as
$ 5.2~{\rm GeV/}c^2 < \mb < 5.3~{\rm GeV/}c^2$ 
and
$-0.12~{\rm GeV} < \dE < 0.25~{\rm GeV}$.
The $\dE$ distribution is modeled with a sum of two (three) Gaussian
functions for $\mb$ ($\dE$).
For the continuum background, 
we use the ARGUS parameterization for $\mb$
and a linear function for $\dE$.
The fit yields $\Nsigomegaks$ $\bz\to\omega\ks$ events
in the signal region
with a statistical significance ($\Sigma$) of 7.3,
where $\Sigma$ is defined as 
$\Sigma \equiv \sqrt{-2\ln(\lzero/\lnsig)}$, and
$\lzero$ and $\lnsig$ denote the maximum likelihoods
of the fits without and with the signal component, respectively.

\subsection{\boldmath $\bz\to\ks\piz$}
Candidate $\ks \to \pip\pim$ decays
are selected with the same criteria as those used for
the $\bz\to\phi\ks$ decay, except that we impose
a more stringent invariant mass requirement;
only pairs of oppositely charged pions that have an invariant mass
within 0.015 GeV/$c^2$
of the nominal $\ks$ mass are used.
The $\piz$ selection criteria are the same as 
those used for the $\bz\to\eta'\ks$ decay.

The $B$ meson signal region is defined as 
$-0.15$ GeV $< \dE <0.1$ GeV
and $5.27~{\rm GeV}/c^2 <\mb<5.29~{\rm GeV}/c^2$.
The dominant background is from continuum.
We use extended modified Fox-Wolfram moments,
which were applied for the selection of
the $\bz\to\piz\piz$ decay~\cite{Abe:2003yy},
to form $\calf$.
We then combine likelihoods for $\calf$ and $\cos\theta_B$
to obtain the event likelihood ratio $\rsigbkg$ for
continuum suppression.
As described below, we include events that do not have
$B$ decay vertex information in our fit
to obtain a better sensitivity for
the $CP$-violation parameter $\cala$.
For events with vertex information,
the high-$\rsigbkg$ region is defined as
$\rsigbkg > 0.74~(0.76)$ for DS-I (DS-II),
and the low-$\rsigbkg$ region as
$0.4 < \rsigbkg \le  0.74~(0.76)$ for DS-I (DS-II).
For events without vertex information,
the high-$\rsigbkg$ region is defined as
$\rsigbkg > 0.78$ 
and the low-$\rsigbkg$ region as
$0.4 < \rsigbkg \le  0.78$
for both DS-I and DS-II.

\begin{figure}
\resizebox{0.3\textwidth}{!}{\includegraphics{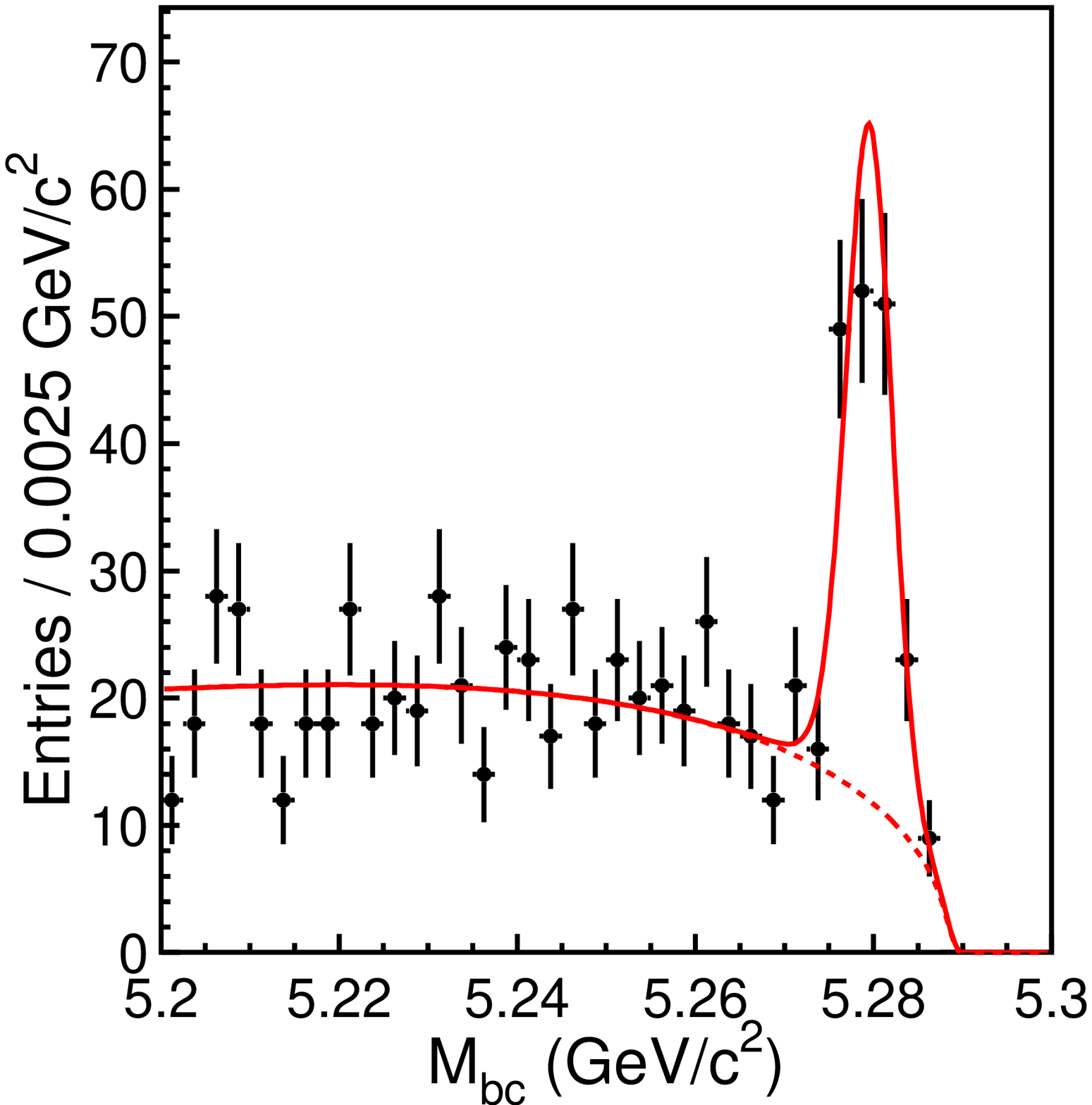}}
\resizebox{0.3\textwidth}{!}{\includegraphics{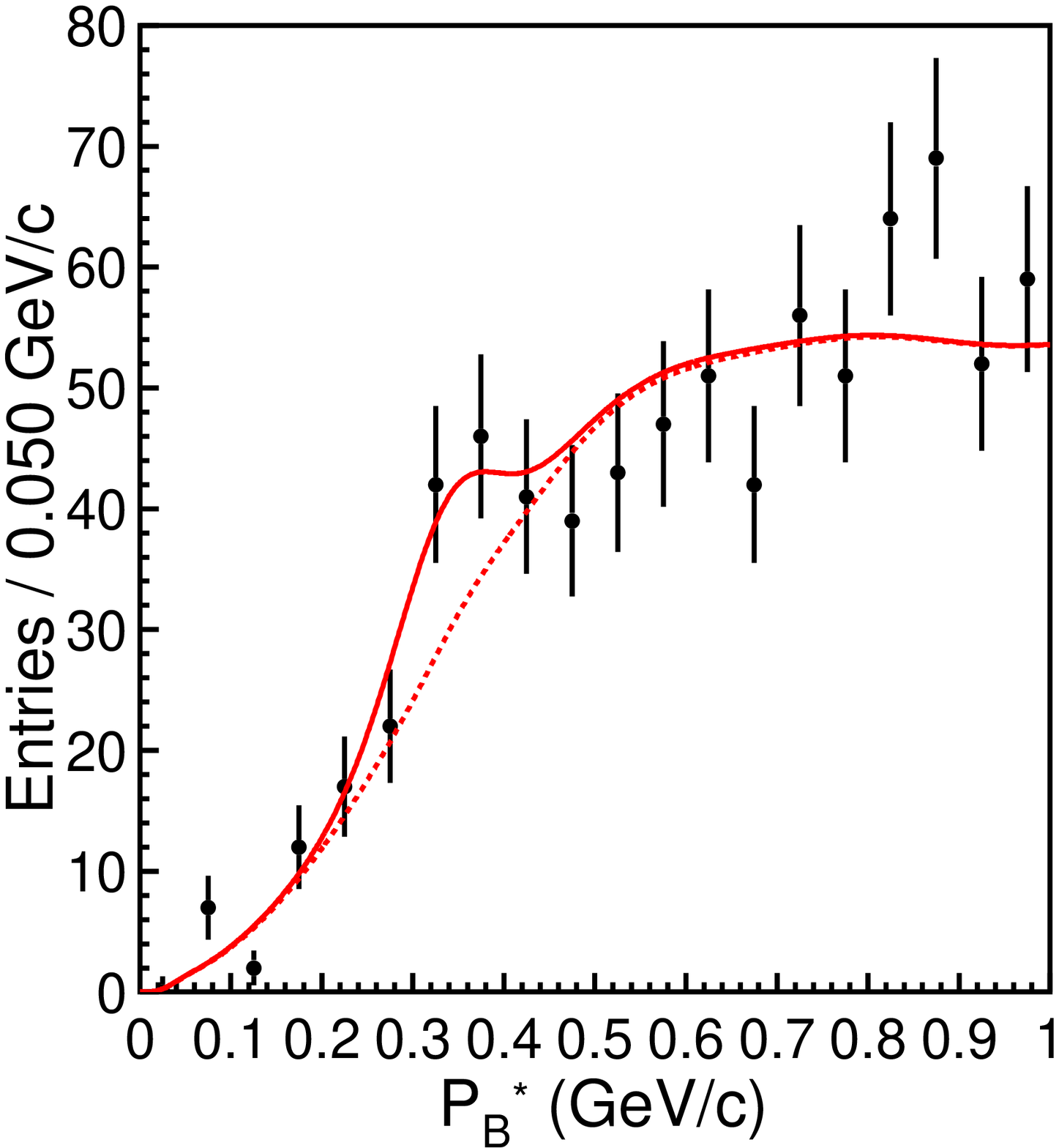}}
\resizebox{0.3\textwidth}{!}{\includegraphics{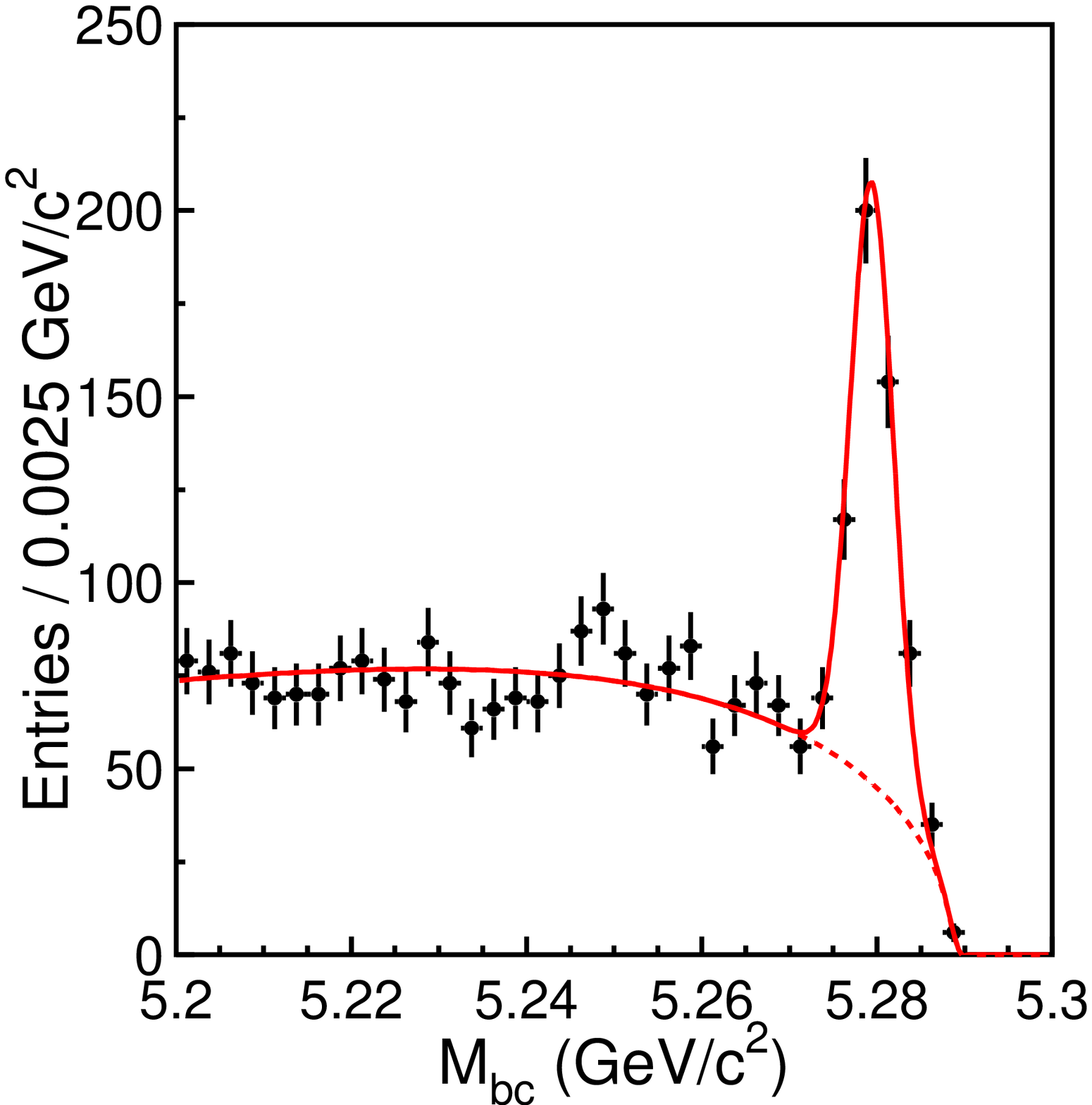}}
\resizebox{0.3\textwidth}{!}{\includegraphics{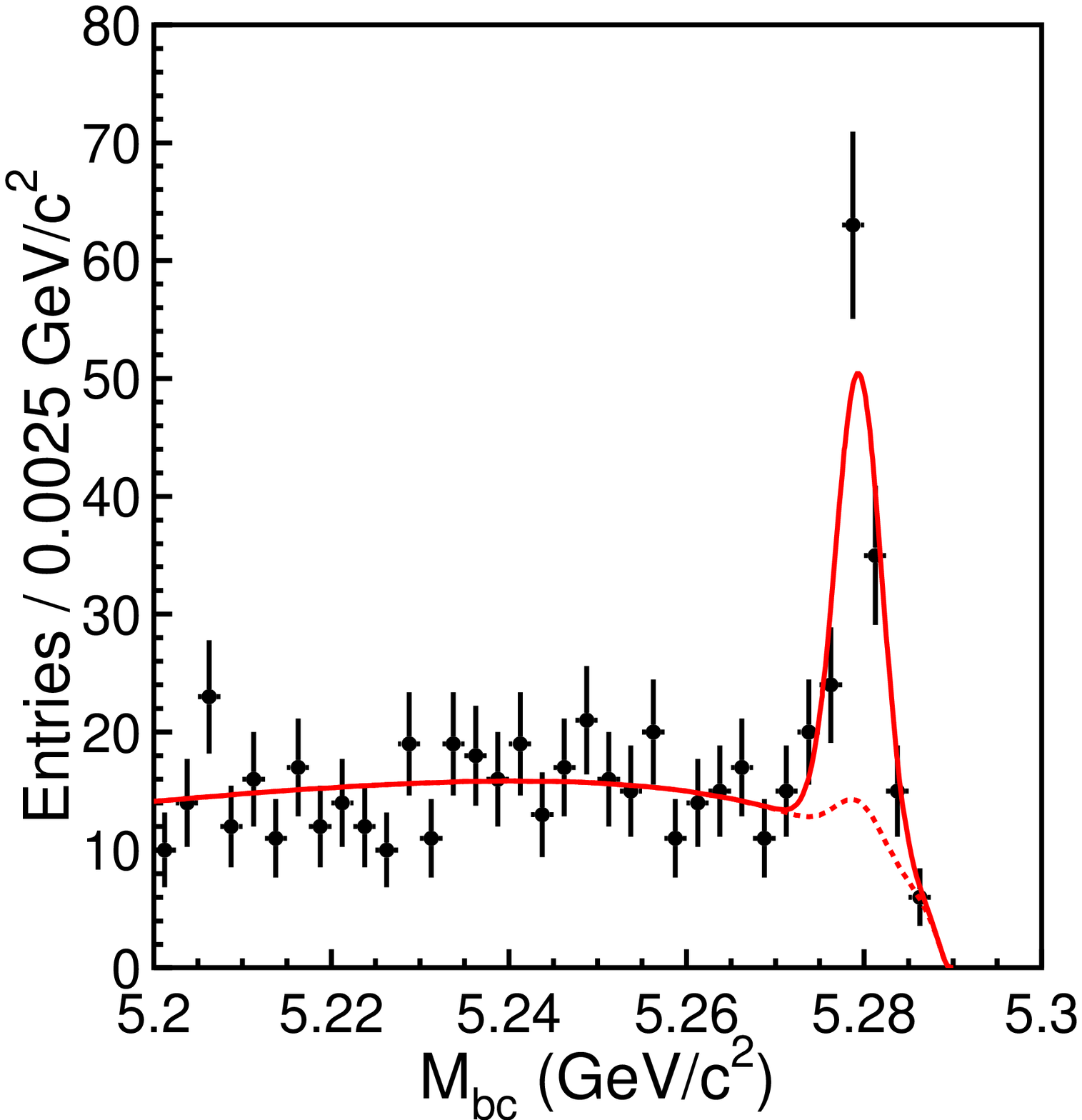}}
\resizebox{0.3\textwidth}{!}{\includegraphics{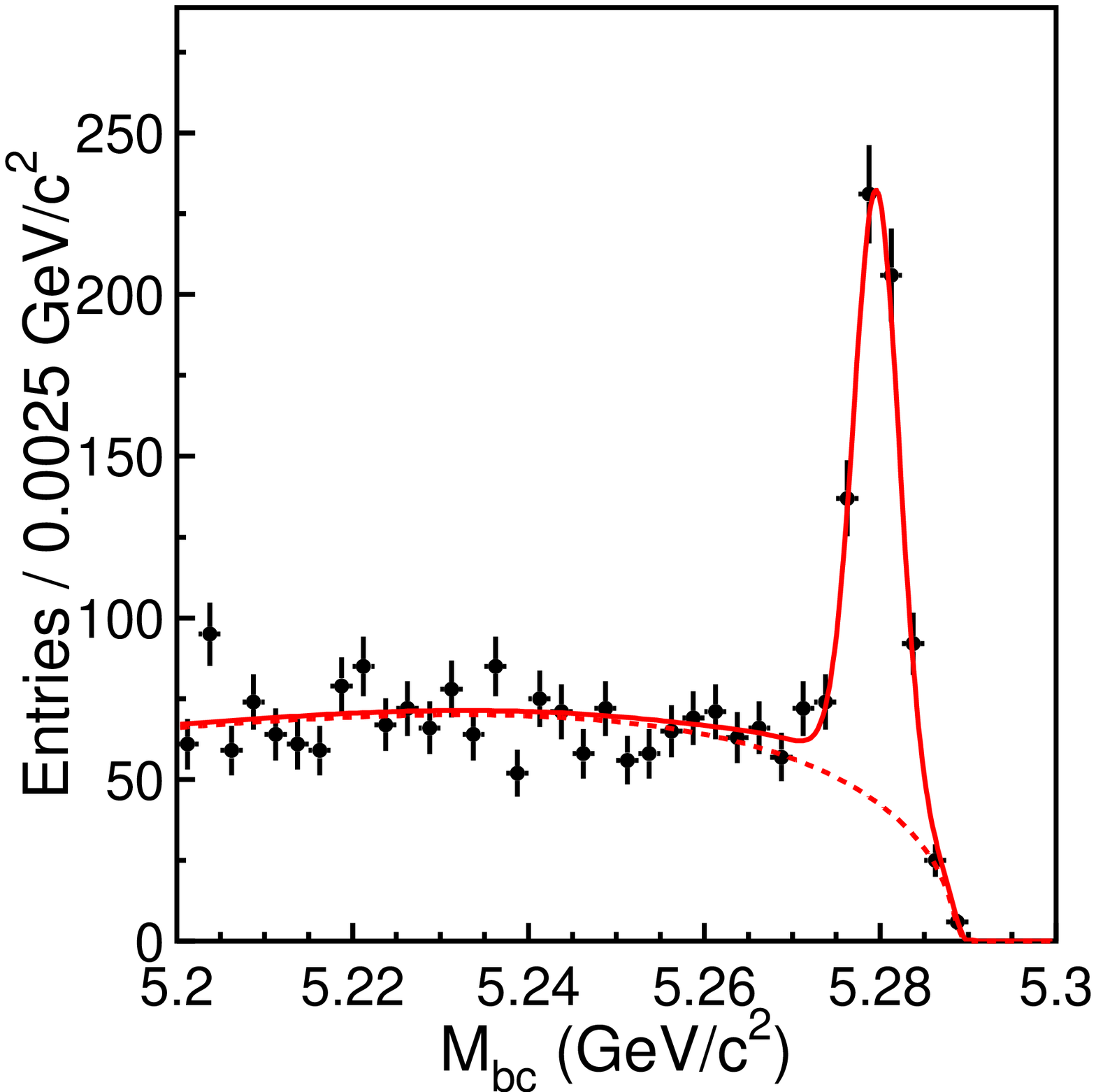}}
\resizebox{0.3\textwidth}{!}{\includegraphics{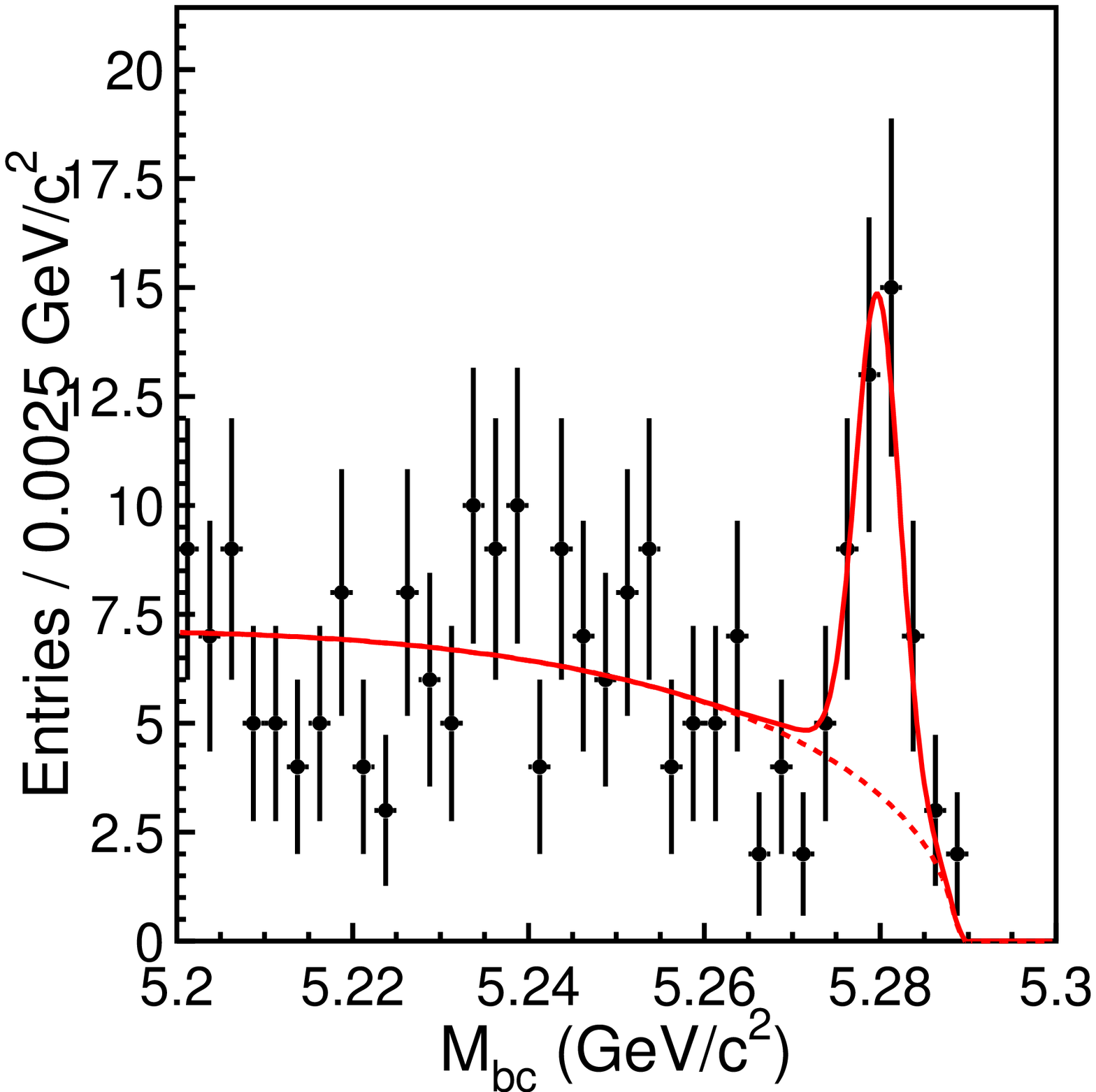}}
\resizebox{0.3\textwidth}{!}{\includegraphics{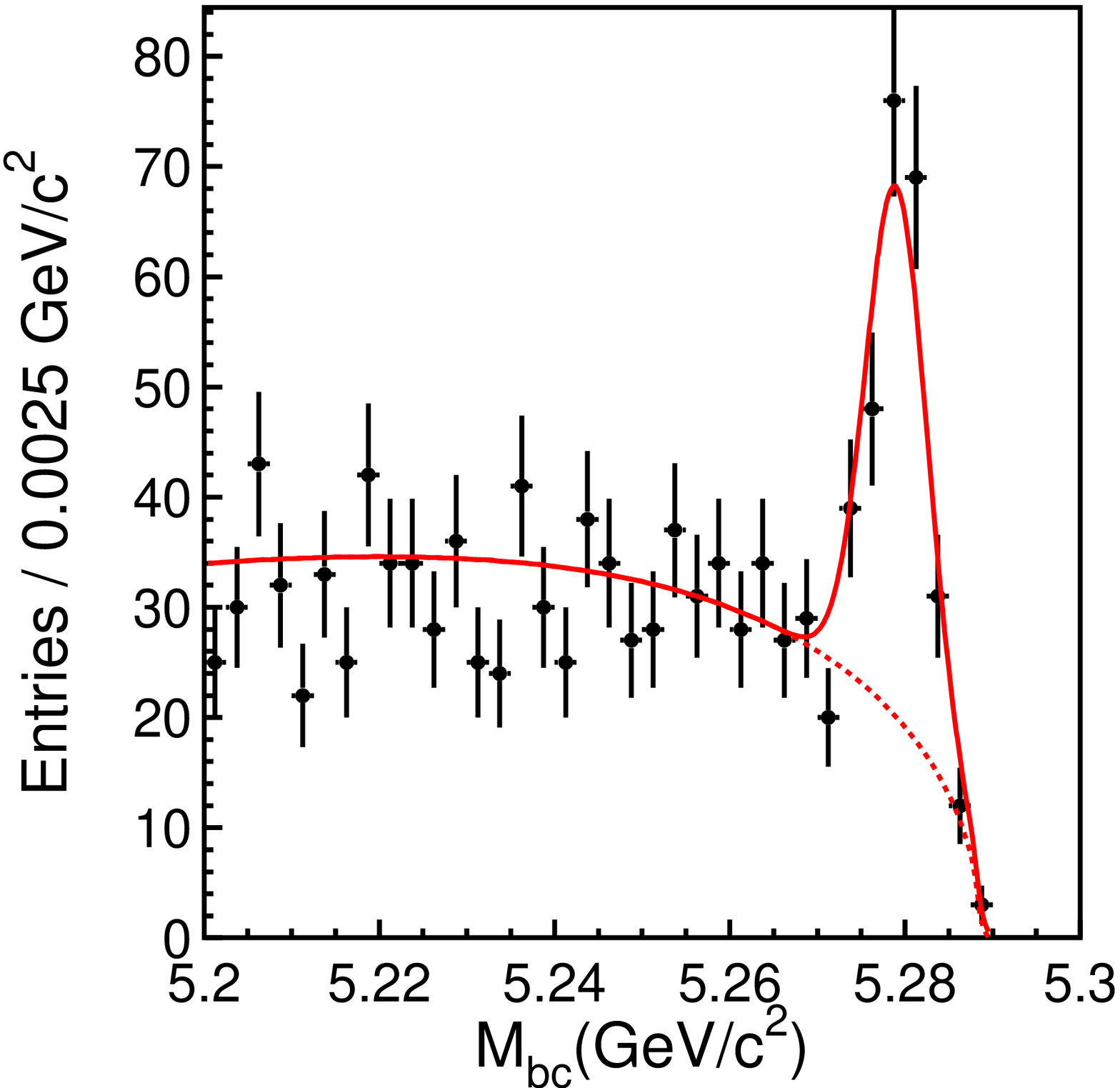}}
\resizebox{0.3\textwidth}{!}{\includegraphics{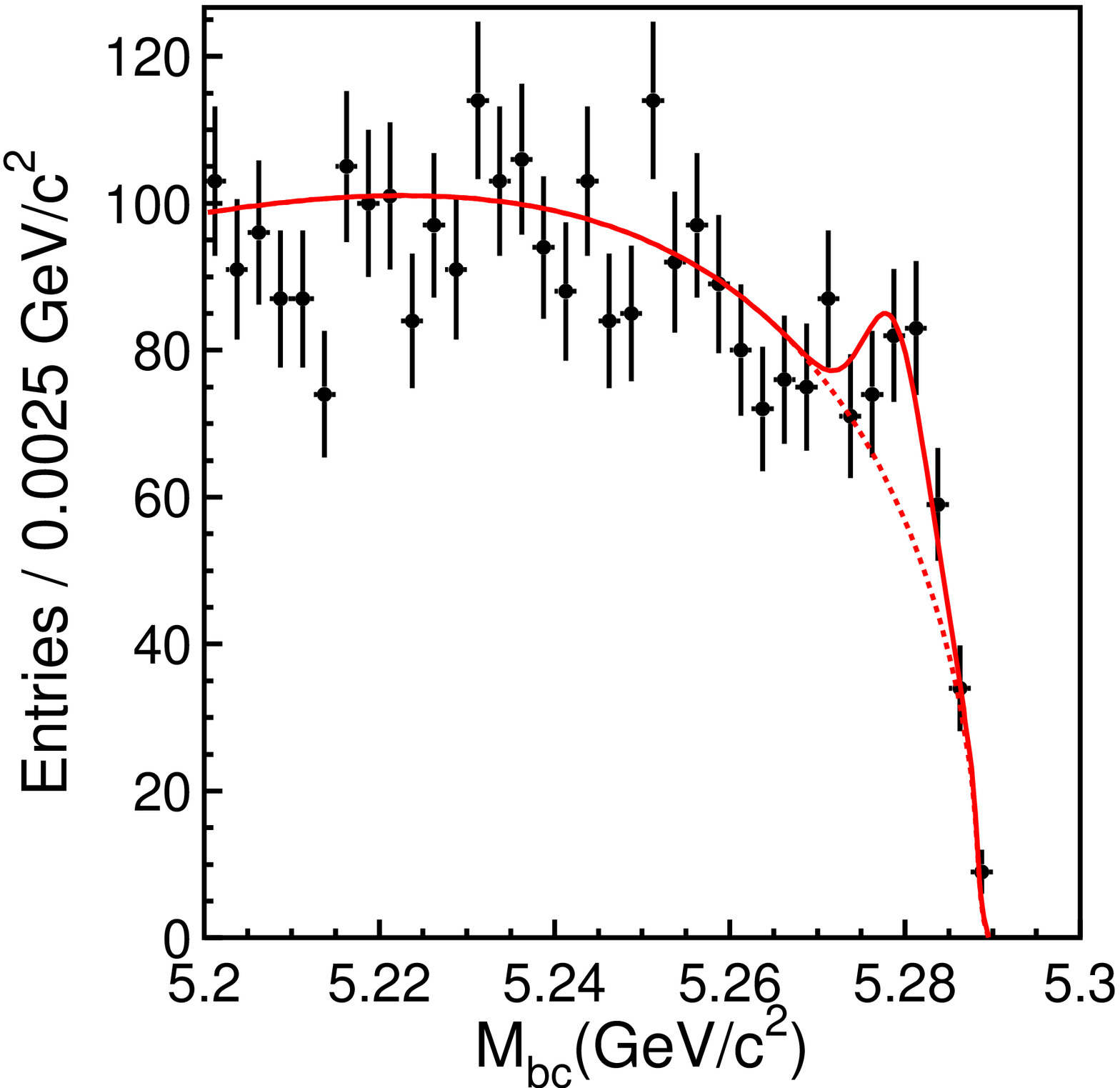}}
\resizebox{0.3\textwidth}{!}{\includegraphics{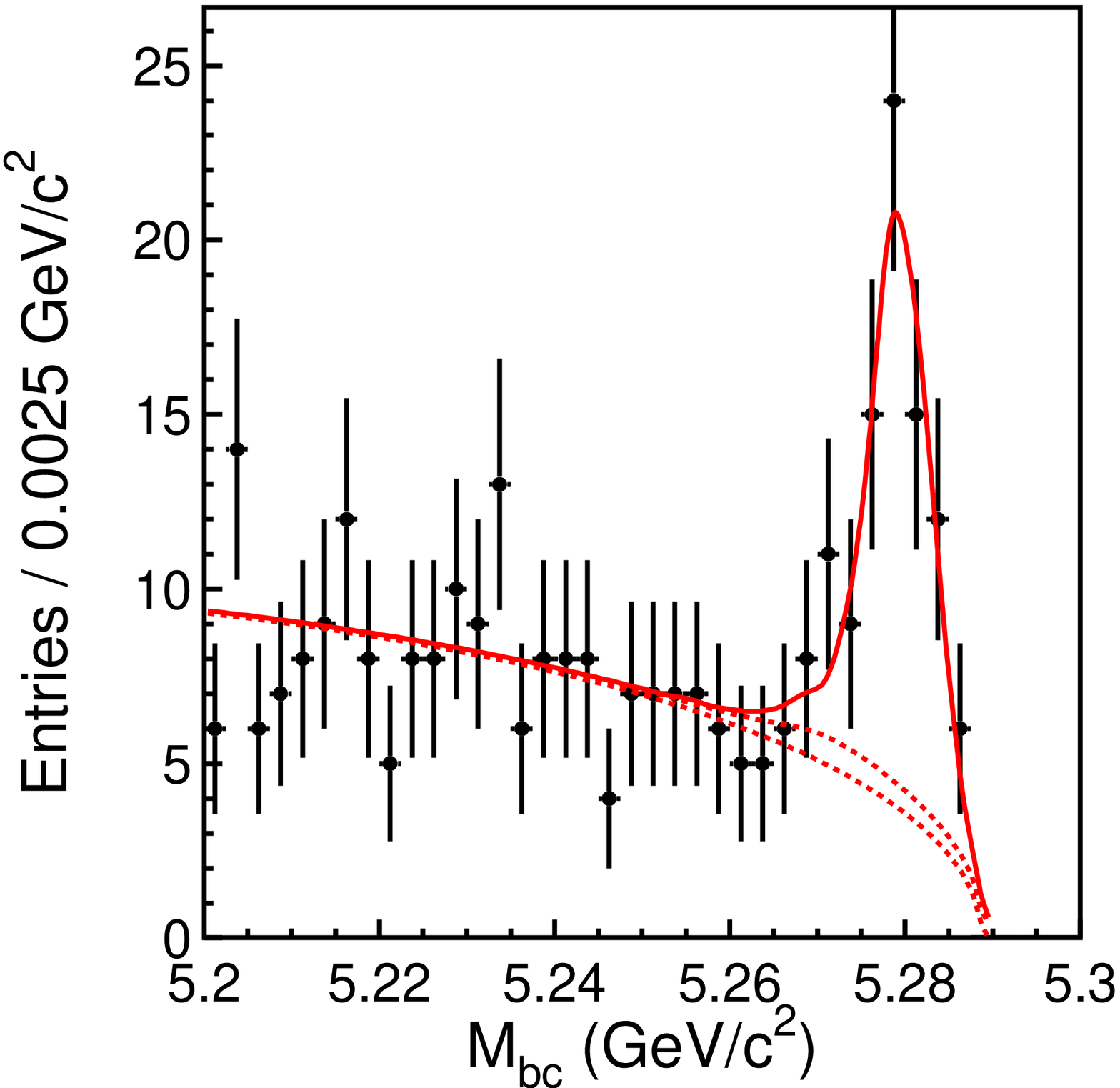}}
\caption{The $\mb$ distributions for
(a) $\bz\to\phi\ks$, (c) $\bz\to\kp\km\ks$, (d) $\bz\to\fzero\ks$,
(e) $\bz\to\eta'\ks$, (f) $\bz\to\omega\ks$, (g) $\bz\to\ks\piz$ 
(high-$\rsigbkg$), 
(h) $\bz\to\ks\piz$ (low-$\rsigbkg$), 
and (i) $\bz\to\kstarz\gamma$ 
within the $\dE$ signal region
and (b) the $p_{B}^{\rm cms}$ distribution for $\bz\to\phi\kl$.
Solid curves show the fit to signal plus background distributions,
and dashed curves show the background contributions.
}\label{fig:mb}
\rput[l](-6.5, 17.5)  {(a)~$\phi\ks$}
\rput[l](-1.4, 17.5)  {(b)~$\phi\kl$}
\rput[l]( 3.65,17.5)  {(c)~$\kp\km\ks$}
\rput[l](-6.5, 12.5)  {(d)~$\fzero\ks$}
\rput[l](-1.4, 12.5)  {(e)~$\eta'\ks$}
\rput[l]( 3.65,12.5)  {(f)~$\omega\ks$}
\rput[l](-6.5,  7.5)  {(g)~$\ks\piz$}
\rput[l](-6.0,  7.0)   {(high-$\rsigbkg$)}
\rput[l](-1.4,  5.0){(h)~$\ks\piz$}
\rput[l](-0.9,  4.5)     {(low-$\rsigbkg$)}
\rput[l]( 3.65, 7.5)  {(i)~$\kstarz\gamma$}
\end{figure}

Figure~\ref{fig:mb}(g) shows
the $\mb$ distribution for the high-$\rsigbkg$ $\bz\to\ks\piz$ candidates
within the $\dE$ signal region
after flavor tagging and before vertex reconstruction.
Also shown in Fig.~\ref{fig:mb}(h) is the $\mb$ distribution
for the low-$\rsigbkg$ $\bz\to\ks\piz$ candidates.
The signal yield is determined
from 
an unbinned
two-dimensional maximum-likelihood fit
to the $\dE$-$\mb$ distribution in the fit region defined as
$ 5.2~{\rm GeV/}c^2 < \mb < 5.29~{\rm GeV/}c^2$ 
and
$-0.2~{\rm GeV} < \dE < 0.5~{\rm GeV}$.
The $\bz\to\ks\piz$ signal distribution is modeled with
a Gaussian function for $\mb$ and with a
Crystal Ball function for $\dE$.
For the continuum background, 
we use the ARGUS parameterization for $\mb$
and a second-order Chebyshev function for $\dE$.
The $B\bbar$ background is negligibly small and its influence is treated 
as a source of systematic uncertainty.
The fits yield $\NsigkspizH$ and $\NsigkspizL$ $\bz\to\ks\piz$ events
in the high-$\rsigbkg$ and 
low-$\rsigbkg$ signal regions, respectively, where the errors are
statistical only.
The same procedure after the vertex reconstruction yields
a total of $77\pm 13$ $\ks\piz$ events.

\subsection{\boldmath $\bz\to\kstarz\gamma~(\kstarz\to\ks\piz)$}
The selection criteria are optimized for the measurement
of time-dependent $CP$ asymmetries and are thus
different from those used
in Belle's previous measurement of
$B\to K^*\gamma$ branching fractions and direct
$CP$ asymmetries~\cite{Nakao:2004th}.
Candidate $\ks \to \pip\pim$ decays
are selected with the same criteria as those used for
the $\bz\to\phi\ks$ decay, except that we impose
a more stringent invariant mass requirement;
only pairs of oppositely charged pions that have an invariant mass
within 0.006 GeV/$c^2$
of the nominal $\ks$ mass are used.
The $\piz$ selection criteria are the same as 
those used for the $\bz\to\eta'\ks$ decay except for
a more stringent $\piz$ momentum requirement,
$\ppizcms > 0.3~{\rm GeV}/c$.
The $\ks\piz$ invariant mass, $M_{\ks\piz}$, is required to be
between 0.6 and 1.8 GeV/$c^2$.

Prompt photons from the $\bz\to\kstarz\gamma$ decay are required
to satisfy $1.4~{\rm GeV} < \egcms < 3.4~{\rm GeV}$,
where $\egcms$ is the photon energy in the cms.
If there is more than one candidate,
the one with the largest $\egcms$ is selected.
For the selected photon,
we require $E_9/E_{25} > 0.95$, 
where $E_9/E_{25}$ is defined in Sec.~\ref{sec:bztophikl}.
Photons for candidate $\piz\to\gamma\gamma$ or 
$\eta\to\gamma\gamma$
decays are not used; we reject photon pairs that satisfy
$\mathcal{L}_{\piz}\ge 0.18$ or $\mathcal{L}_{\eta}\ge 0.18$,
where $\mathcal{L}_{\piz(\eta)}$ is a 
$\piz$ ($\eta$) likelihood described in detail
elsewhere~\cite{Koppenburg:2004fz}.
The polar angle of the
photon direction in the laboratory frame is required to be between
$33^\circ$ and $128^\circ$ for DS-I, 
while no requirement is imposed for DS-II 
as the material within the acceptance of the ECL is much reduced
for this dataset.

Candidate $\kstarp\to\ks\pip$ decays are also selected using a similar
procedure to reconstruct the decay
$\bz\to\kstarp\gamma$.
Candidate $\bz\to\kstarz\gamma~(\kstarz\to\ks\piz)$ and
$\bp\to\kstarp\gamma~(\kstarp\to\ks\pip)$ decays are selected simultaneously;
we allow only one candidate for each event.
The best candidate selection is based on the event likelihood ratio
$\rsigbkg$ that is obtained by
combining $\calf$, which uses
the extended modified Fox-Wolfram moments as
discriminating variables, 
with $\cos\theta_H$ defined as the
angle between the $\bz$ meson momentum and the daughter
$\ks$ momentum in the $\kstarz$ meson rest frame.
We select the candidate with the largest $\rsigbkg$.

The signal region for the $\bz\to\kstarz\gamma~(\kstarz\to\ks\piz)$ decay
is defined as 
$-0.2$ GeV $< \dE <0.1$ GeV,
$5.27~{\rm GeV}/c^2 <\mb<5.29~{\rm GeV}/c^2$ and
$0.8{\rm GeV}/c^2 < M_{\ks\piz} < 1.0~{\rm GeV}/c^2$.
We require $\rsigbkg > 0.5$ to reduce the continuum background.

We use events outside the signal region 
as well as a large MC sample to study the background components.
The dominant background is from continuum.
Background contributions from $B$ decays 
are significantly smaller than those from continuum,
and are dominated by
cross-feed from $\bp\to\kstarp\gamma$ decays,
other radiative $B$ decays and charmless $B$ decays.
Background from other $B\bbar$ decays is found to be negligible.

Figure~\ref{fig:mb}(i) shows
the $\mb$ distribution for the reconstructed $\bz\to\kstarz\gamma$ candidates
within the $\dE$ signal region
after flavor tagging and vertex reconstruction.
The signal yield is determined
from an unbinned 
two-dimensional maximum-likelihood fit
to the $\dE$-$\mb$ distribution in the fit region defined as
$ 5.20~{\rm GeV/}c^2 < \mb < 5.29~{\rm GeV/}c^2$ 
and
$-0.5~{\rm GeV} < \dE < 0.5~{\rm GeV}$.
The $\bz\to\kstarz\gamma$ signal distribution is represented by
a smoothed histogram obtained from MC simulation that
accounts for the correlation between $\mb$ and $\dE$.
The background from $B$ decays is also modeled with
a smoothed histogram obtained from MC events;
its normalization is a free parameter in
the fit. For the continuum background, 
we use the ARGUS parameterization for $\mb$
and a second-order Chebyshev function for $\dE$.
The fit yields $\Nsigkstarzgm$ $\bz\to\kstarz\gamma~(\kstarz\to\ks\piz)$
events, where the error is statistical only.
For reference, we also measure the signal
before vertex reconstruction and obtain
$132\pm 14$ events. 

\subsection{Flavor Tagging}
\label{sec:flavor tagging}
The $b$-flavor of the accompanying $B$ meson is identified
from inclusive properties of particles
that are not associated with the reconstructed $\bz \to \fCP$ 
decay. We use the same procedure that is used for the
$\sinbb$ measurement~\cite{bib:BELLE-CONF-0436}.
The algorithm for flavor tagging is described in detail
elsewhere~\cite{bib:fbtg_nim}.
We use two parameters, $\fq$ and $r$, to represent the tagging information.
The first, $\fq$, is already defined in Eq.~(\ref{eq:psig}).
The parameter $r$ is an event-by-event,
MC-determined flavor-tagging dilution factor
that ranges from $r=0$ for no flavor
discrimination to $r=1$ for unambiguous flavor assignment.
It is used only to sort data into six $r$ intervals.
The wrong tag fractions for the six $r$ intervals, 
$w_l~(l=1,6)$, and differences 
between $\bz$ and $\bzb$ decays, $\dwl$,
are determined from the data;
we use the same values
that were used for the $\sin 2\phi_1$ measurement~\cite{bib:BELLE-CONF-0436}
for DS-I.
Wrong tag fractions for DS-II are separately obtained 
with the same procedure; we find that the values for DS-II,
which are listed in Table~\ref{tab:wtag},
are slightly smaller than those for DS-I.
The total effective tagging efficiency for DS-II
is determined to be
$\eeff \equiv \sum_{l=1}^6 \epsilon_l(1-2w_l)^2 = \efftot$,
where $\epsilon_l$ is the event fraction for each $r$ interval
determined from the $\jpsi\ks$ data and listed in
Table~\ref{tab:wtag}.
The error includes both statistical and systematic uncertainties.

\subsection{Vertex Reconstruction}
The vertex position for the $\fCP$ decay 
is reconstructed using charged tracks that have enough SVD hits.
A constraint on the IP is also used with the selected tracks;
the IP profile is convolved with finite $B$ flight length in the plane
perpendicular to the $z$ axis.
The pions from $\ks$ decays are not used
except for the analysis of
$\bz\to\kstarz\gamma$ and $\bz\to\ks\piz$ decays.
The vertex
for $\bz\to\kstarz\gamma$ and $\bz\to\ks\piz$ decays is
reconstructed using
the $\ks$ trajectory and the IP constraint, where
both pions from the $\ks$ decay are required to
have enough SVD hits to reconstruct a vertex. 
The reconstruction efficiency depends both on
the $\ks$ momentum and on the SVD geometry;
efficiencies with SVD-II are higher than
those with SVD-I
because of the larger outer radius and the additional layer.

The $\ftag$ vertex determination with SVD-I
remains unchanged from the
previous publication~\cite{Abe:2003yt},
and is described in detail elsewhere~\cite{bib:resol_nim};
to minimize the effect of long-lived particles, 
secondary vertices from charmed hadrons and a small fraction of
poorly reconstructed tracks, we adopt an iterative procedure
in which the track that gives the largest contribution to the
vertex $\chi^2$ is removed at each step 
until a good $\chi^2$ is obtained.

For SVD-II, we find that
the same vertex reconstruction algorithm results in
a larger outlier fraction when only
one track remains after the iteration procedure.
Therefore, in this case, we repeat the
iteration procedure with
a more stringent requirement on the SVD-II hit pattern.
The resulting outlier fraction is comparable to
that for SVD-I, while 
the inefficiency caused by this change is small (2.5\%).

\subsection{Summary of Signal Yields}
The signal yields for $\bz\to\fCP$ decays, $\nsig$,
after flavor tagging and vertex reconstruction
(before the vertex reconstruction for the
decay $\bz\to\ks\piz$)
are summarized in Table~\ref{tab:num}.
The signal purities are also listed in the table.

\section{Results of {\boldmath $CP$} Asymmetry Measurements}
We determine $\cals$ and $\cala$ for each mode by performing an unbinned
maximum-likelihood fit to the observed $\Dt$ distribution.
The probability density function (PDF) expected for the signal
distribution, ${\cal P}_{\rm sig}(\Dt;\cals,\cala,\fq,w_l,\dwl)$, 
is given by Eq.~(\ref{eq:psig}) incorporating
the effect of incorrect flavor assignment. The distribution is
convolved with the
proper-time interval resolution function 
$R_{\rm sig}(\Dt)$,
which takes into account the finite vertex resolution. 

For the decays $\bz\to\phi\ks$, $\kp\km\ks$, $\phi\kl$, $\fzero\ks$, $\eta'\ks$
and $\omega\ks$,
we use flavor-specific $B$ decays governed by
semileptonic or hadronic $b\to c$ transitions
to determine the resolution function.
We perform a simultaneous multi-parameter fit to these high-statistics
control samples to obtain the resolution function parameters,
wrong-tag fractions (Section~\ref{sec:flavor tagging}), 
$\dmd$, $\taubp$ and $\taubz$.
We use the same resolution function used for
the $\sinbb$ measurement for DS-I~\cite{bib:BELLE-CONF-0436}.
For DS-II, the following modifications are introduced:
a sum of two Gaussian functions is used
to model the resolution of the $\fCP$ vertex
while a single Gaussian function is used for DS-I;
a sum of two Gaussian functions is used to model
the resolution of the tag-side vertex
obtained with one track and the IP constraint,
while a single Gaussian function is used for DS-I.
These modifications are needed to account for
differences for SVD-I and SVD-II, as well as
different background conditions in DS-I and DS-II.
We test the new resolution parameterization using
MC events on which we overlay
beam-related background taken from data.
A fit to the MC sample yields correct values
for all parameters. 
With the multi-parameter fit to data, we find that
the standard deviation of the main Gaussian component
of the resolution function
is reduced from 78 $\mu$m to 55 $\mu$m, which
is consistent with our expectation from
the improved impact parameter resolution of SVD-II~\cite{Ushiroda}.
The same fit also yields
$\taubz = 1.518\pm 0.012$ ps,
$\taubp = 1.652\pm 0.014$ ps
and 
$\dmd = 0.516\pm 0.007~{\rm ps}^{-1}$.
The results are consistent with those obtained with 
DS-I~\cite{bib:BELLE-CONF-0436}
and also with the world average values~\cite{bib:PDG2004}.
Thus we conclude that the resolution of SVD-II is well understood.

For the decays $\bz\to\ks\piz$ and $\kstarz\gamma~(\kstarz\to\ks\piz)$,
we use the resolution function described above
with additional parameters that rescale vertex
errors. The rescaling parameters depend on
the detector configuration (SVD-I or SVD-II), 
SVD hit patterns of charged pions from the $\ks$ decay,
and $\ks$ decay vertex position in the plane 
perpendicular to the beam axis.
These parameters are determined from a fit to the 
$\Dt$ distribution of $\bz\to\jpsi\ks$ data.
Here the $\ks$ and the IP constraint are used for the
vertex reconstruction, the $\bz$ lifetime is
fixed at the world average value, and $b$-flavor tagging
information is not used so that the 
expected PDF is
an exponential function convolved with
the resolution function.

We check the resulting resolution function
by also reconstructing the vertex with
leptons from $\jpsi$ decays and the IP constraint.
We find that the distribution of the
distance between the vertex positions obtained with
the two methods is well represented by
the obtained resolution function convolved with
the well-known resolution for the $\jpsi$ vertex.
Finally, we also perform a fit to the $\bz\to\jpsi\ks$ sample
with $b$-flavor information and obtain
$\cals_{\jpsi\ks} = +0.68\pm 0.10$(stat) and
$\cala_{\jpsi\ks} = +0.02\pm 0.04$(stat), which
are in good agreement with the world average values.
Thus, we conclude that 
the vertex resolution for the $\bz\to\ks\piz$ and
$\bz\to\kstarz\gamma~(\kstarz\to\ks\piz)$ decays
is well understood.

We determine the following likelihood value for each
event:
\begin{eqnarray}
P_i
&=& (1-\fol)\int \biggl[
\fsig{\cal P}_{\rm sig}(\Dt')R_{\rm sig}(\Dt_i-\Dt') \nonumber \\
&+&(1-\fsig){\cal P}_{\rm bkg}(\Dt')R_{\rm bkg}(\Dt_i-\Dt')\biggr]
d(\Dt')  \nonumber \\
&+&\fol P_{\rm ol}(\Dt_i) 
\label{eq:likelihood}
\end{eqnarray}
where $P_{\rm ol}(\Dt)$ is a broad Gaussian function that represents
an outlier component with a small fraction $\fol$.
The signal probability $\fsig$ depends on the $r$ region and
is calculated on an event-by-event basis
as a function of $\pbstar$ for the $\bz\to\phi\kl$ decay and
as a function of $\dE$ and $\mb$ for the other modes.
A PDF for background events, ${\cal P}_{\rm bkg}(\Dt)$,
is modeled as a sum of exponential and prompt components, and
is convolved with a sum of two Gaussians $R_{\rm bkg}$.
All parameters in ${\cal P}_{\rm bkg} (\Dt)$
and $R_{\rm bkg}$ are determined by the fit to the $\Dt$ distribution of a 
background-enhanced control sample~\cite{bib:BBbg}; i.e. events 
outside of the
$\dE$-$\mb$ signal region.
We fix $\tau_\bz$ and $\dmd$ at
their world-average values~\cite{bib:PDG2004}.
In order to reduce the statistical error on $\cala$,
we include events without vertex information
in the analysis of $\bz\to\ks\piz$.
The likelihood value in this case is obtained by integrating 
Eq.~(\ref{eq:likelihood}) over $\Dt_i$.

The only free parameters in the final fit
are $\cals$ and $\cala$, which are determined by maximizing the
likelihood function
$L = \prod_iP_i(\Dt_i;\cals,\cala)$
where the product is over all events.
Since $\cala$ for the decay $\bz\to\kstarz\gamma$ is
well measured in the $\bz\to\kstarz\gamma~(\kstarz\to\kp\pim)$ mode
and is consistent with zero, we fix $\cala$ at zero and
perform a fit to the $\bz\to\kstarz\gamma~(\kstarz\to\ks\piz)$ sample
with $\cals$ as the only free parameter.

\begin{figure}
\resizebox{!}{0.32\textwidth}{\includegraphics{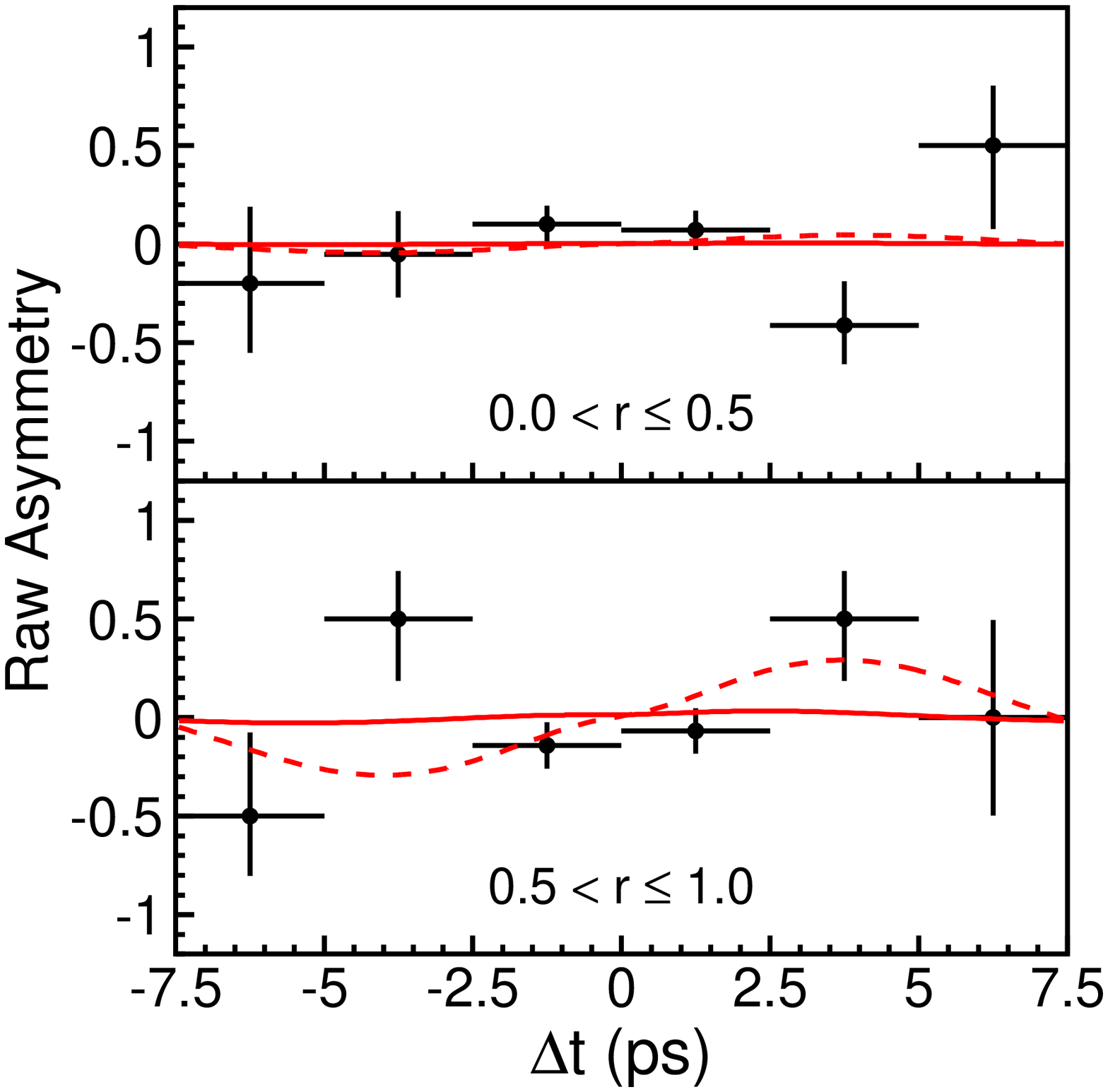}} 
\resizebox{!}{0.32\textwidth}{\includegraphics{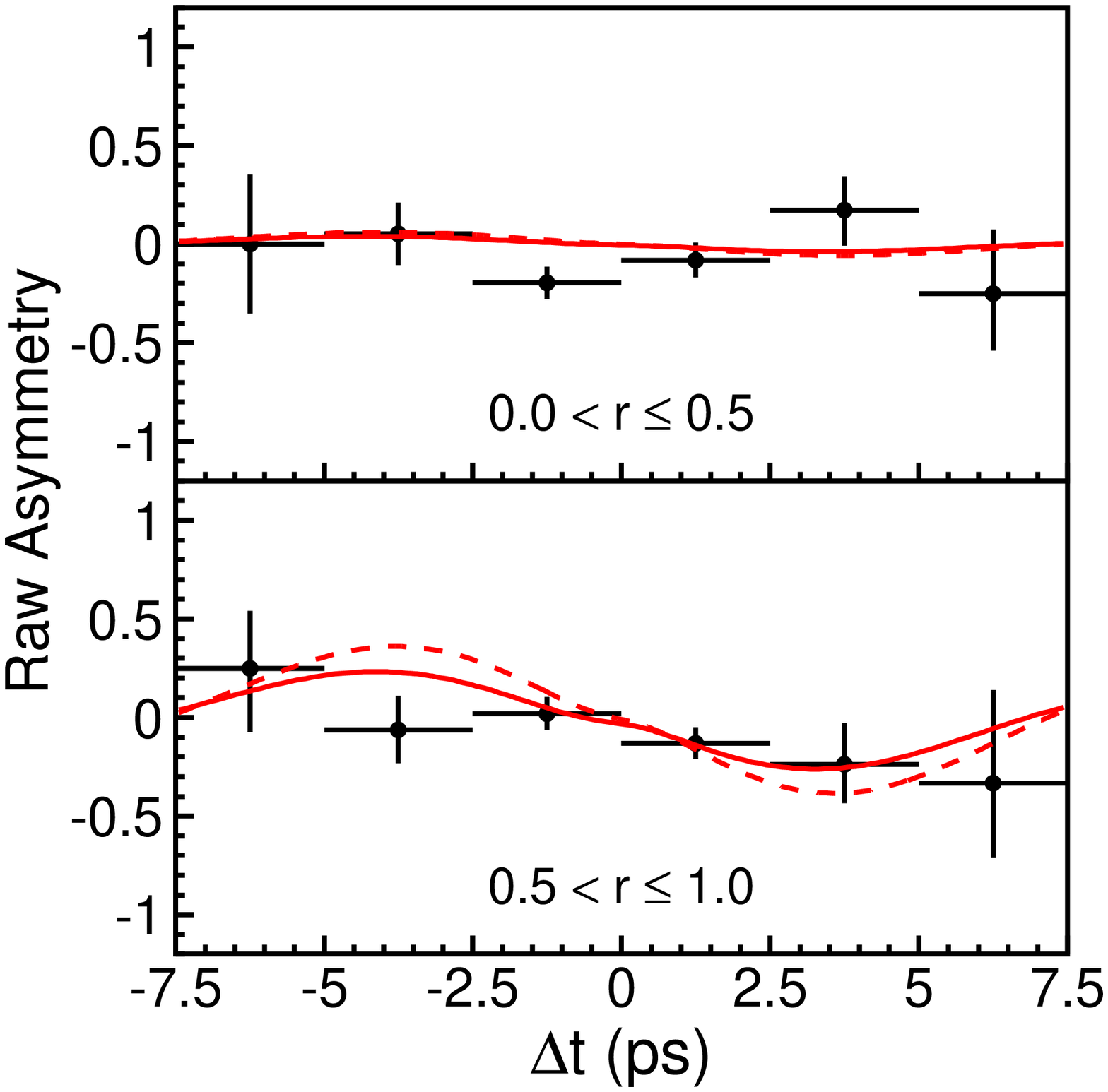}} 
\resizebox{!}{0.32\textwidth}{\includegraphics{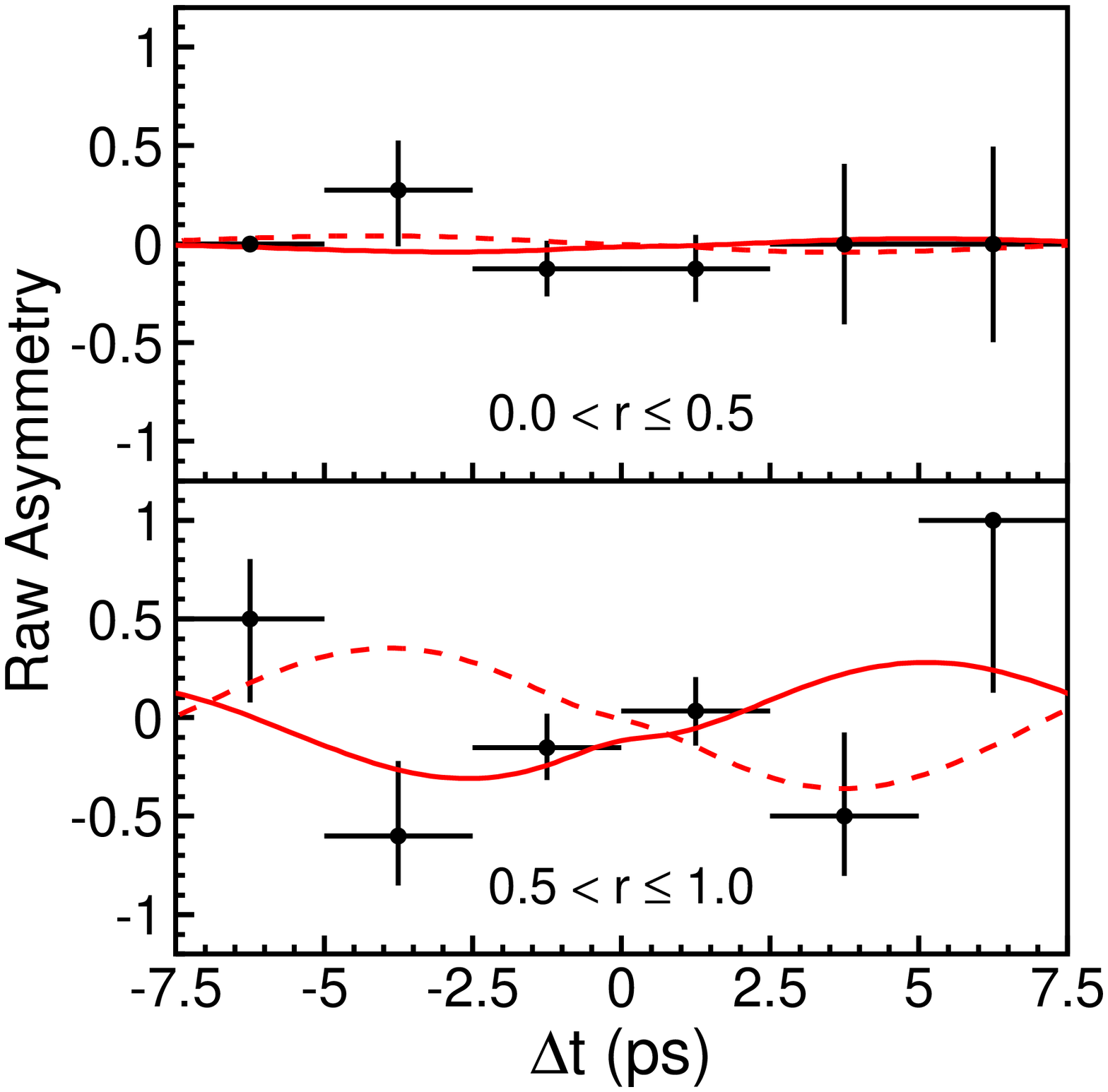}} 
\resizebox{!}{0.32\textwidth}{\includegraphics{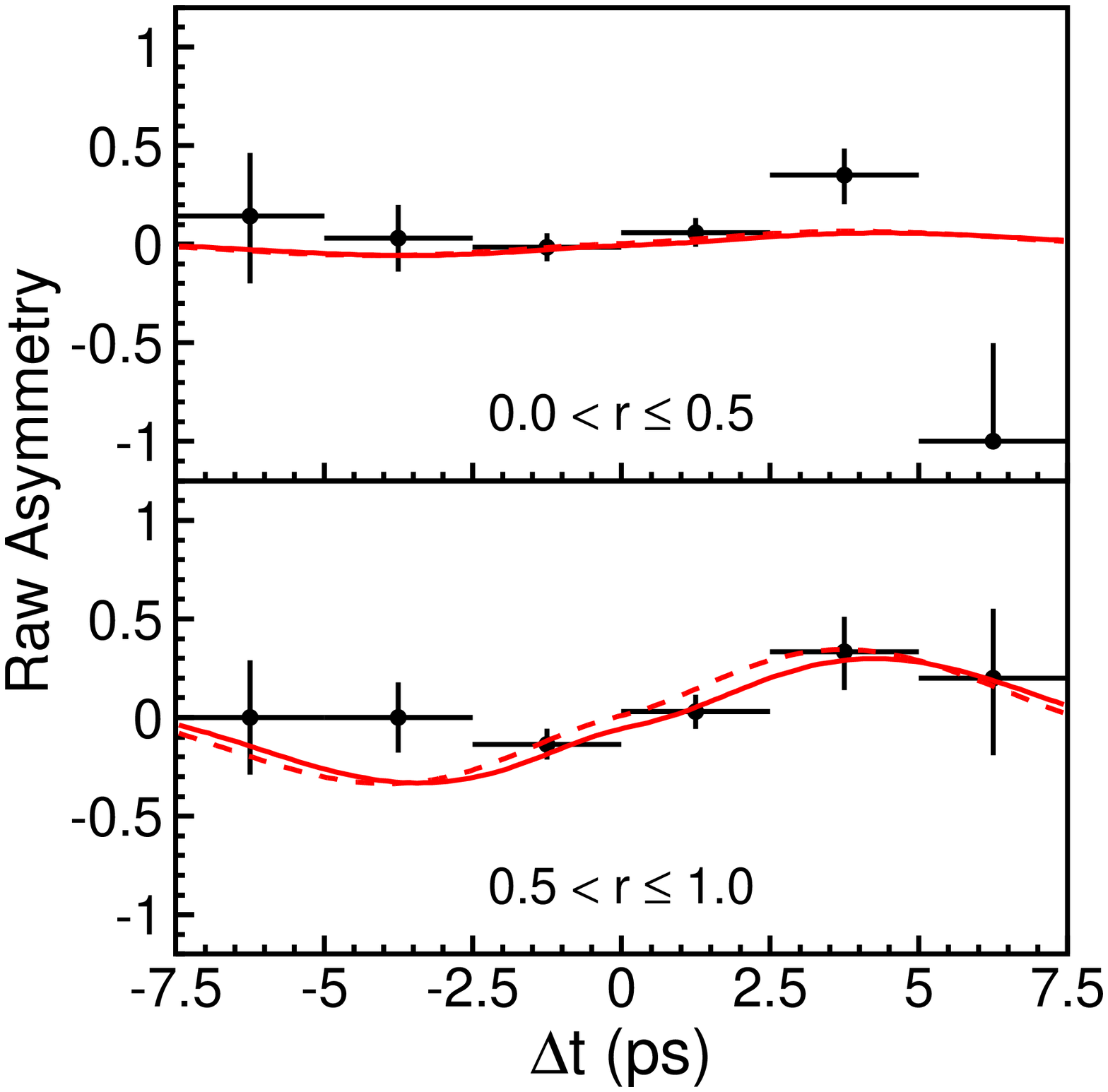}} 
\resizebox{!}{0.32\textwidth}{\includegraphics{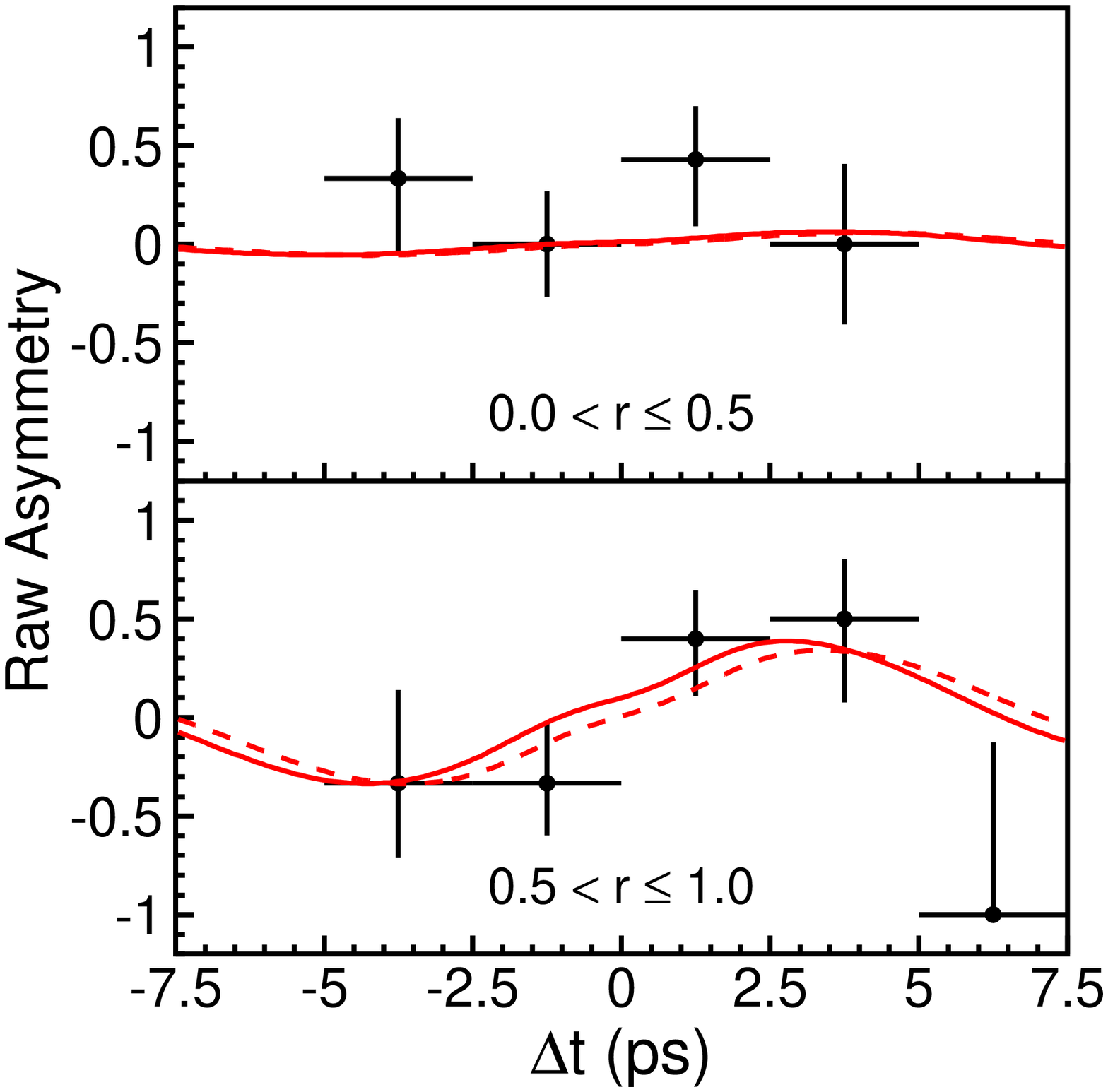}} 
\resizebox{!}{0.32\textwidth}{\includegraphics{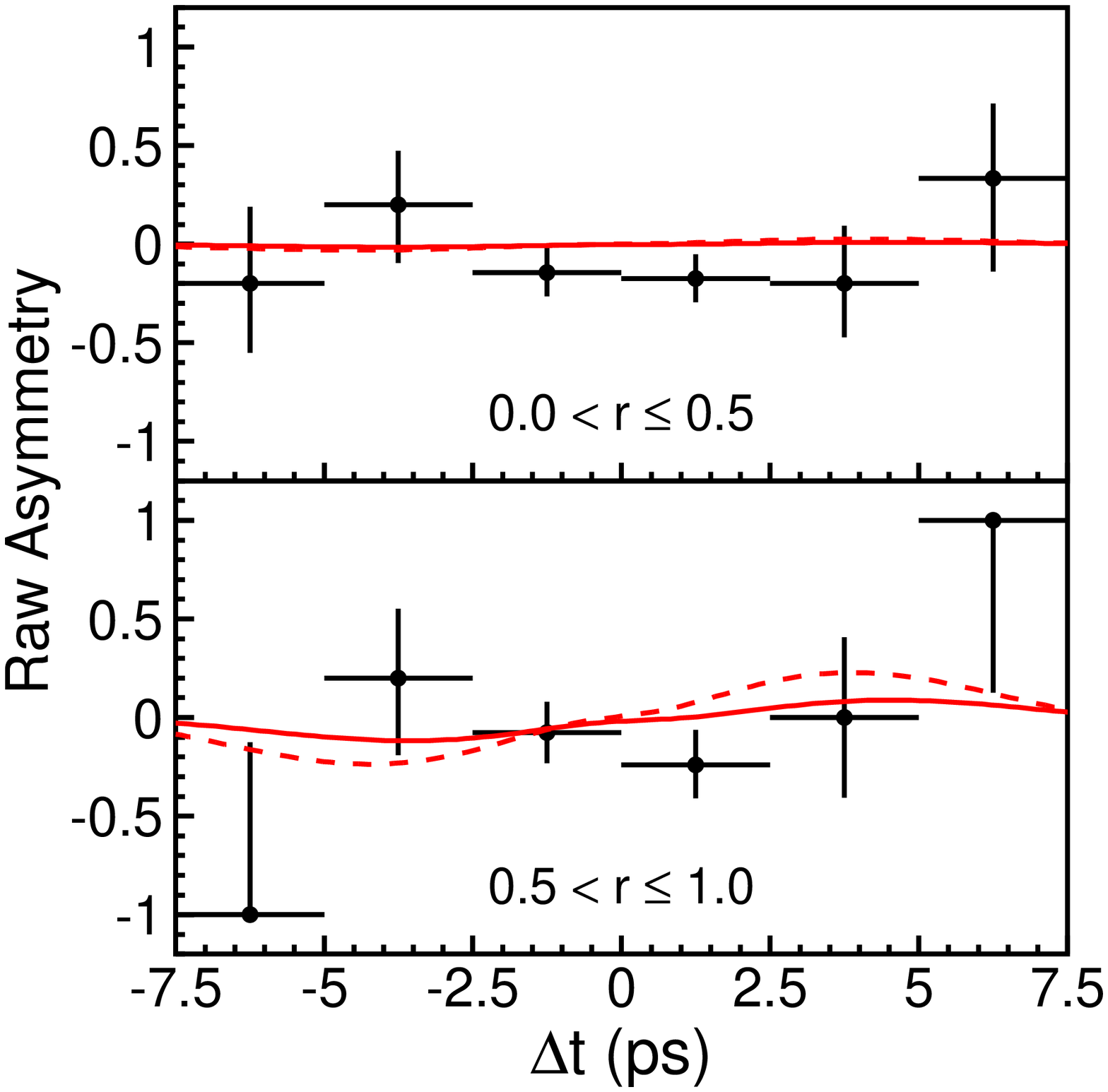}} 
\resizebox{!}{0.32\textwidth}{\includegraphics{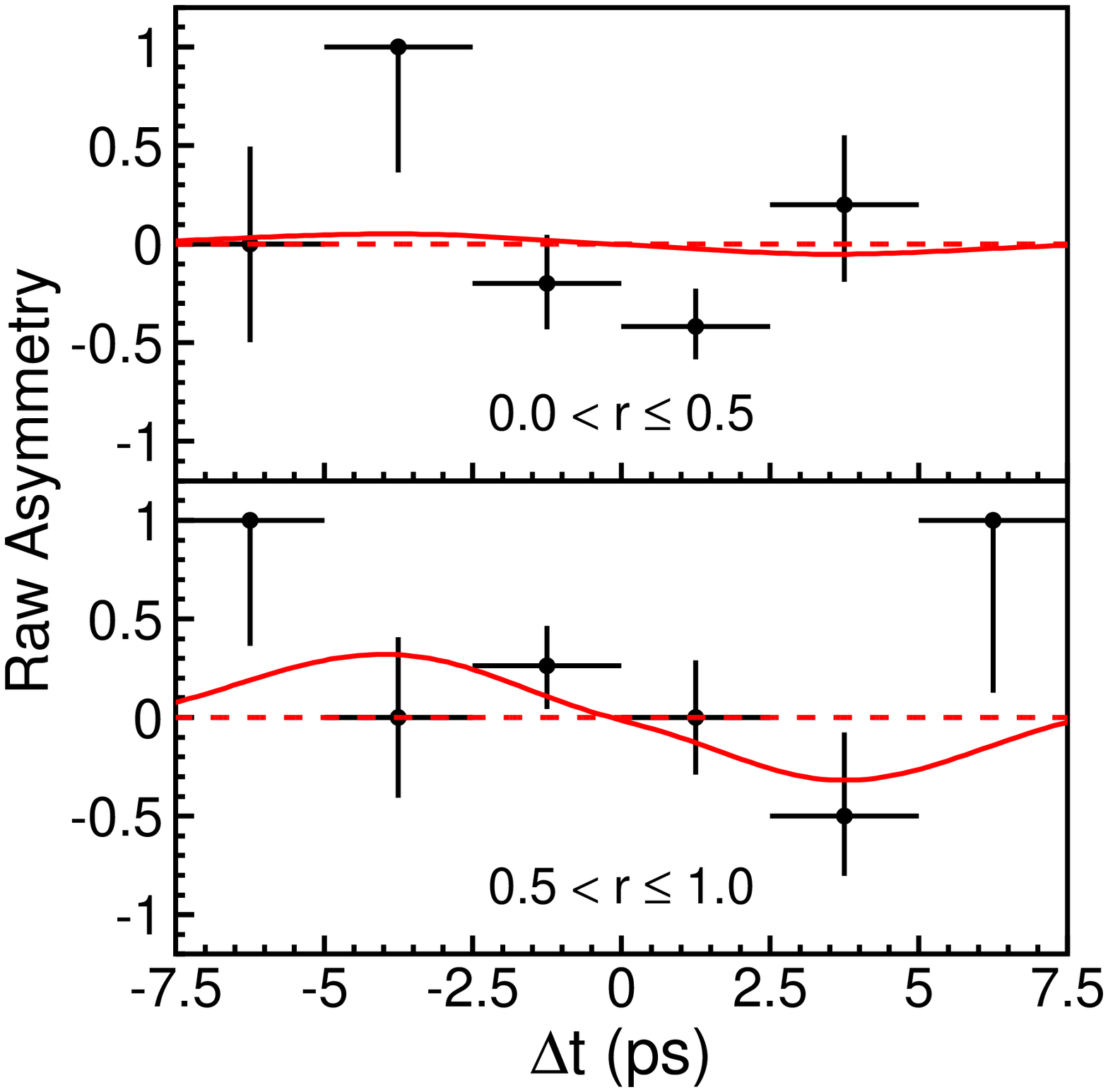}} 
\caption{
The asymmetry, $A$, in each $\Dt$ bin with $0 < r \le 0.5$ (top)
and with $0.5 < r \le 1.0$ (bottom) for 
(a) $\bz\to\phi\kz$, 
(b) $\bz\to\kp\km\ks$, (c) $\bz\to\fzero\ks$,
(d) $\bz\to\eta'\ks$, (e) $\bz\to\omega\ks$, 
(f) $\bz\to\ks\piz$ (for the high-$\rsigbkg$ region only), 
and (g) $\bz\to\kstarz\gamma$.
The solid curves show the result of the 
unbinned maximum-likelihood fit.
The dashed curves show the SM expectation with $\sin2\phi_1$ = +0.73 
($\cals = 0$ for $\bz\to\kstarz\gamma$) and $\cala$ = 0.}
\rput[l](-6.9, 18.5)  {(a)~$\phi\kz$}
\rput[l](-1.5, 18.5)  {(b)~$\kp\km\ks$}
\rput[l]( 3.85,18.5)  {(c)~$\fzero\ks$}
\rput[l](-6.9, 13.2)  {(d)~$\eta'\ks$}
\rput[l](-1.5, 13.2)  {(e)~$\omega\ks$}
\rput[l]( 3.75,13.2)  {(f)~$\ks\piz$ (high-$\rsigbkg$)}
\rput[l]( 0.5,  7.9)  {(g)~$\kstarz\gamma$}
\label{fig:asym}
\end{figure}

Table \ref{tab:result} summarizes
the fit results of $\cals$ and $\cala$.
We define the raw asymmetry in each $\Dt$ bin by
$(N_{q=+1}-N_{q=-1})/(N_{q=+1}+N_{q=-1})$,
where $N_{q=+1(-1)}$ is the number of 
observed candidates with $q=+1(-1)$~\cite{footnote:phikz}.
Figures~\ref{fig:asym}(a-g) show the raw asymmetries in two regions of the flavor-tagging
parameter $r$. While the numbers of events in the two regions are similar,
the effective tagging efficiency is much larger 
and the background dilution is smaller in the region $0.5 < r \le 1.0$.
Note that these projections onto the $\Delta t$ axis do not take into
account event-by-event information (such as the signal fraction, the
wrong tag fraction and the vertex resolution), which is used in the
unbinned maximum-likelihood fit.

Tables~\ref{tab:ssyserr} and \ref{tab:asyserr}
list the systematic errors on $\cals$ and $\cala$, respectively.
The total systematic errors are obtained
by adding each contribution in quadrature,
and are much smaller than the statistical errors for all modes.

To determine the systematic error that arises from
uncertainties in the vertex reconstruction,
the track and vertex selection criteria
are varied to search for possible systematic biases.
Small biases in the $\Dz$ measurement 
are observed in $e^+e^-\to\mu^+\mu^-$ and other control
samples. Systematic errors 
are estimated by applying special correction functions
to account for the observed biases, repeating
the fit, and comparing the obtained values with the nominal results.
The systematic error due to the IP constraint 
in the vertex reconstruction is estimated by
varying ($\pm10 \mu$m) the smearing used to account for the
$B$ flight length.
Systematic errors due to imperfect SVD alignment
are determined
from MC samples that have artificial mis-alignment effects
to reproduce impact-parameter resolutions observed in data.

Systematic errors due to uncertainties in the wrong tag
fractions are studied by varying
the wrong tag fraction individually for each $r$ region.
Systematic errors due to uncertainties in the resolution function
are also estimated by varying each resolution parameter obtained from
data (MC) by $\pm 1\sigma$ ($\pm 2\sigma$), repeating the fit
and adding each variation in quadrature.
Each physics parameter such as $\taubz$ and $\dmd$
is also varied by its error.
A possible fit bias is examined by fitting a large number of MC events.

Systematic errors from uncertainties in the background fractions
and in the background $\Dt$ shape
are estimated by varying each background parameter obtained
from data (MC) by $\pm 1\sigma$ ($\pm 2\sigma$).
Uncertainties in the background $B$ decay model
are also considered for
the $\bz\to\kstarz\gamma~(\kstarz\to\ks\piz)$ mode;
we compare different theoretical models for radiative $B$ decays
and take the largest variation as the systematic error.

Additional sources of systematic errors are 
considered for $B$ decay backgrounds
that are neglected in the PDF.
We consider uncertainties both in their fractions
and $CP$ asymmetries; for modes that have
non-vanishing $CP$ asymmetries, we conservatively
vary the $CP$-violation parameters within the
physical region and take the largest variation
as the systematic error.
The effect of backgrounds from $\kp\km\ks$ and 
$\fzero\ks~(\fzero\to\kp\km)$
in the $\bz\to\phi\ks$ sample is considered.
Uncertainties from
$B\to\phi K^*$ and other rare $B$ decay backgrounds
in the $\bz\to\phi\kl$ sample
are also taken into account.
The peaking background fraction in the $\bz\to\fzero\ks$ sample 
depends on the PDF used in the fit to the 
$\pip\pim$ invariant mass distribution, which
ignores possible interference between
resonant and non-resonant amplitudes.
We perform a fit to the $\pip\pim$ distribution
of a MC sample generated with
interfering
amplitudes and phases for $B\to K\pi\pi$ decays measured
from data~\cite{Garmash:2003er}. The observed difference in the
signal yield from the true value is taken into account in the 
systematic error determination.
We also repeat the fit to the $\Dt$ distribution
ignoring the contribution of the peaking background.
The differences in $\cals$ and $\cala$ from
our nominal results are included in the systematic error.

Finally, we investigate the effects of interference between
CKM-favored and CKM-suppressed $B\to D$ transitions in
the $\ftag$ final state~\cite{Long:2003wq}.
A small correction to the PDF for the signal distribution
arises from the interference.
We estimate the size of the correction using the $\bzdslnu$ 
sample. We then generate MC pseudo-experiments
and make an ensemble test to obtain systematic biases
in $\cals$ and $\cala$. We find that
the effect on $\cals$ is negligibly small, while
a possible shift in $\cala$ is sizable.

Various crosschecks of the measurement are performed.
We reconstruct charged $B$ meson decays
that are the counterparts of the $\bz\to\fCP$ decays
and apply the same fit procedure.
All results for the $\cals$ term are consistent with no 
$CP$ asymmetry, as expected. 
Lifetime measurements are also performed for 
the $\fCP$ modes and the corresponding charged $B$ decay modes.
The fits yield
$\taubz$ and $\taubp$ values consistent with the world-average values.
MC pseudo-experiments are generated for each decay mode to
perform ensemble tests.
We find that the statistical errors obtained
in our measurements are all consistent
with the expectations from the ensemble tests.

For the $\bz\to\phi\kz$ decay,
a fit to DS-I alone yields
$\cals = -0.68\pm 0.46$(stat) and 
$\cala = -0.02 \pm 0.28$(stat),
while a fit to DS-II alone yields
$\cals = +0.78\pm 0.45$(stat) and $\cala = +0.17 \pm 0.33$(stat).
Note that the results for DS-I
differ from our previously published results
$\cals = \SphiksResPrv$ and $\cala = \AphiksResPrv$~\cite{Abe:2003yt},
as decays $\bz\to\phi\kl$ and $\phi\ks~(\ks\to\piz\piz)$ are
included in this analysis.
From MC pseudo-experiments, 
the probability that the difference between $\cals$ values in
DS-I and DS-II is larger than the observed difference (1.46)
is estimated to be 4.5\%.
A $\sinbb$ measurement with DS-II is performed
using $\bz\to\jpsi\ks~(\ks\to \pip\pim~{\rm or}~\piz\piz)$ and
$\bz\to\jpsi\kl$ decays as a crosscheck.
Applying the same procedure to both DS-I and DS-II,
we obtain
$\cals_{\jpsi\kz} = +0.696\pm0.061$(stat) 
and
$\cala_{\jpsi\kz} = +0.011\pm0.043$(stat)
for DS-I, and
$\cals_{\jpsi\kz} = +0.629\pm0.069$(stat) 
and
$\cala_{\jpsi\kz} = +0.035\pm0.044$(stat)
for DS-II.
The results are in good agreement with each other, and
are also consistent with SM expectations.
As all the other checks mentioned above also yield
results consistent with expectations,
we conclude that the difference in $\cals_{\phi\kz}$
between the two datasets is due to a statistical fluctuation.

A fit to the $\bz\to\kstarz\gamma~(\kstarz\to\ks\piz)$ sample with
both $\cala$ and $\cals$ as free parameters yields 
$\cals = \SkstarzgmVal\SkstarzgmStat$(stat),
which is identical with the result of the one parameter fit, 
and
$\cala = \AkstarzgmVal\pm\AkstarzgmStat$(stat),
which is consistent with zero.

\begin{figure}
\resizebox{0.9\textwidth}{!}{\includegraphics{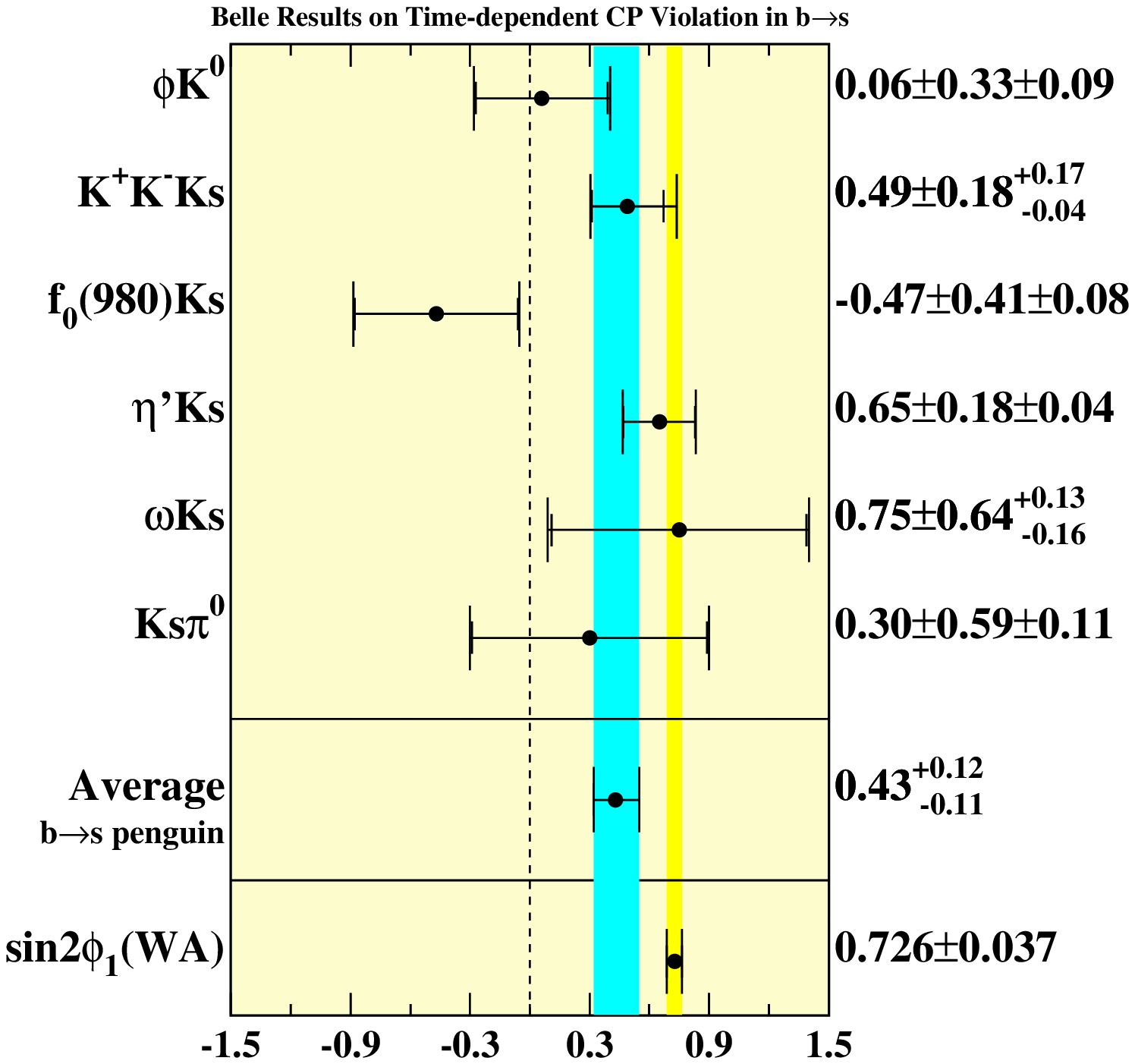}} 
\caption{Summary of $\sinbb$ measurements performed with
$\bz$ decay governed by the $b\to s \overline{q}q$ transition. 
The world-average $\sinbb$ value obtained from $\bz\to\jpsi\kz$
and other related decay modes governed by the $b\to c\overline{c}s$
transition~\cite{bib:HFAG} is also shown as the SM reference.}
\label{fig:avg}
\end{figure}

As discussed in Section~\ref{sec:introduction},
to a good approximation,
the SM predicts $\cals = -\xi_f\sin 2\phi_1$
for the $\bz\to\phi\kz$, $\kp\km\ks$, $\fzero\ks$, $\eta'\ks$,
$\omega\ks$ and $\ks\piz$ decays. 
Figure~\ref{fig:avg} summarizes 
the $\sinbb$ determination based on our $\cals$ measurements
for these decays.
For each mode, the first error shown in the figure
is statistical and the second error
is systematic. 
For the $\bz\to\kp\km\ks$ decay,
an additional systematic error that arises
from the uncertainty of the $CP$-even component
fraction ($^{+0.17}_{-0.00}$) is added in quadrature.
We obtain $\sinbb = \SbsqqResult$ as a weighted average,
where the error includes both statistical and systematic errors.
The result differs from the SM expectation by
2.4 standard deviations.

\section{Summary}
We have performed improved measurements of 
$CP$-violation parameters for $\bz \to \phi \kz$ (including
both $\phi\ks$ and $\phi\kl$), $K^+K^-\ks$ and $\eta'\ks$ decays, 
and new measurements for
$\bz\to\fzero\ks$, $\omega\ks$ and $\ks\piz$ decays.
These charmless decays
are dominated by $b\to s$ flavor-changing neutral currents
and are sensitive to possible new $CP$-violating phases.
We have also measured the time-dependent $CP$ asymmetry
in the decay $\kstarz\gamma~(\kstarz\to\ks\piz)$, which
is also sensitive to physics beyond the SM.
The results for each individual decay mode are consistent with
the SM expectation within two standard deviations
except for the $\bz\to\fzero\ks$ decay.
The combined result for
the $\bz\to\phi\kz$, $\kp\km\ks$, $\fzero\ks$, $\eta'\ks$,
$\omega\ks$ and $\ks\piz$ decays
differs from the SM expectation by 2.4 standard deviations.
Measurements with a much larger data sample are required to 
conclusively establish the existence of a new $CP$-violating phase
beyond the SM.

\input{ichep04ack.tex}


\clearpage
\newpage

\begin{table*}
  \caption{The event fractions $\epsilon_l$,
    wrong-tag fractions $w_l$, wrong-tag fraction differences $\dwl$,
    and average effective tagging efficiencies
    $\eeff^l = \epsilon_l(1-2w_l)^2$ for each $r$ interval for the DS-II.
    The errors for $w_l$ and $\dwl$
    include both statistical and systematic uncertainties.
    The event fractions are obtained from $\jpsi\ks$ data.}
  \begin{ruledtabular}
    \begin{tabular}{ccclll}
      $l$ & $r$ interval & $\epsilon_l$ &\multicolumn{1}{c}{$w_l$} 
          & \multicolumn{1}{c}{$\dwl$}  &\multicolumn{1}{c}{$\eeff^l$} \\
      \hline
 1 & 0.000 -- 0.250 & $0.397\pm 0.015$ & $0.464\pm0.007$ &$+0.010\pm0.007$ &$0.002\pm0.001$ \\
 2 & 0.250 -- 0.500 & $0.146\pm 0.009$ & $0.321\pm0.008$ &$-0.022\pm0.010$ &$0.019\pm0.002$ \\
 3 & 0.500 -- 0.625 & $0.108\pm 0.008$ & $0.224\pm0.011$ &$+0.031\pm0.011$ &$0.033\pm0.004$ \\
 4 & 0.625 -- 0.750 & $0.107\pm 0.008$ & $0.157\pm0.010$ &$+0.002\pm0.011$ &$0.051\pm0.005$ \\
 5 & 0.750 -- 0.875 & $0.098\pm 0.007$ & $0.109\pm0.009$ &$-0.028\pm0.011$ &$0.060\pm0.005$ \\
 6 & 0.875 -- 1.000 & $0.144\pm 0.009$ & $0.016\pm0.005$ &$+0.007\pm0.007$ &$0.135\pm0.009$ \\
    \end{tabular}
  \end{ruledtabular}
\label{tab:wtag} 
\end{table*}
\begin{table}
\caption{
The estimated signal purity and
the signal yield $\nsig$ in the signal region for each $\fCP$ mode
that is used to measure $CP$ asymmetries.
The result for the $\bz\to\ks\piz$ decay is obtained
with the sample after flavor tagging but
before vertex reconstruction as
events that do not have vertex information are also
used to extract the direct $CP$ violation parameter $\cala$.
The results for the other decays are obtained
after flavor tagging and vertex reconstruction.}
\label{tab:num}
\begin{ruledtabular}
\begin{tabular}{llllr}
\multicolumn{1}{c}
{Mode}    &         &$\xi_f$        
                           &\multicolumn{1}{c}{purity} 
                                          & \multicolumn{1}{c}{$\nsig$} \\
\hline
$\phi\ks$         & & $-1$ & $\Pphiks$    & $\Nsigphiks$ \\
$\phi\kl$         & & $+1$ & $\Pphikl$    & $\Nsigphikl$ \\
$K^+K^-\ks$       & & $+1$ & $\Pkpkmks$   & $\Nsigkpkmks$ \\
$\fzero\ks$       & & $+1$ & $\Pfzeroks$  & $\Nsigfzeroks$ \\
$\eta'\ks$        & & $-1$ & $\Petapks$   & $\Nsigetapks$ \\
$\omega\ks$       & & $-1$ & $\Pomegaks$  & $\Nsigomegaks$ \\
$\ks\piz$ &(high-$\rsigbkg$)& $-1$ & $\PkspizH$   & $\NsigkspizH$ \\
          &(low-$\rsigbkg$) & $-1$ & $\PkspizL$   & $\NsigkspizL$ \\
$\kstarz(\ks\piz)
         \gamma$  & &      & $\Pkstarzgm$ & $\Nsigkstarzgm$ \\
\end{tabular}
\end{ruledtabular}
\end{table}
\begin{table}
\caption{Results of the fits to the $\Dt$ distributions.
The first error is statistical and the second
error is systematic. We combine
$\bz\to\phi\ks$ and $\bz\to\phi\kl$ decays
to obtain $\cals_{\phi\kz}$ and $\cala_{\phi\kz}$.
}
\label{tab:result}
\begin{ruledtabular}
\begin{tabular}{lrll}
\multicolumn{1}{c}{Mode} &  
SM expectation for $\cals$ &
\multicolumn{1}{c}{$\cals$} & 
\multicolumn{1}{c}{$\cala$} \\
\hline
$\phi\kz$   & $+\sinbb$   & $\SphikzResult$    & $\AphikzResult$    \\
$K^+K^-\ks$ & $-\sinbb$   & $\SkpkmksResult$   & $\AkpkmksResult$   \\
$\fzero\ks$ & $-\sinbb$   & $\SfzeroksResult$  & $\AfzeroksResult$  \\
$\eta'\ks$  & $+\sinbb$   & $\SetapksResult$   & $\AetapksResult$   \\
$\omega\ks$ & $+\sinbb$   & $\SomegaksResult$  & $\AomegaksResult$  \\
$\ks\piz$   & $+\sinbb$   & $\SkspizResult$    & $\AkspizResult$    \\
$\kstarz\gamma~(\kstarz\to\ks\piz)$ 
            & $-2(m_s/m_b)\sinbb$ 
                          & $\SkstarzgmResult$ & \multicolumn{1}{c}{-} \\
\end{tabular}
\end{ruledtabular}
\end{table}
\begin{table*}
  \caption{Summary of the systematic errors on $\cals$.}
  \begin{ruledtabular}
    \begin{tabular}{lrrrrrrr}
           & $\phi\kz$ 
                    &$\kp\km\ks$ 
                           &$\fzero\ks$ 
                                   &$\eta'\ks$
                                            &$\omega\ks$
                                                  &$\kstarz\gamma$
                                                              &$\ks\piz$\\
\hline
Vertex 
reconstruction & 0.01 & 0.01  & 0.02  & 0.01  & 0.01   & 0.06  & 0.02 \\
Flavor tagging & 0.01 &$<0.01$& 0.01  & 0.01  & 0.04   & 0.02  & 0.01 \\
Resolution 
function       & 0.04 & 0.03  & 0.03  & 0.03  & 0.07   & 0.05  & 0.05 \\
Physics 
parameter   & $<0.01$ &$<0.01$& 0.01  &$<0.01$& 0.01   & 0.01  & 0.02 \\
Possible 
fit bias       & 0.01 & 0.01  & 0.03  & 0.01&$^{+0.01}
                                              _{-0.10}$& 0.03  & 0.03 \\
Background 
fraction  & $^{+0.08}
             _{-0.06}$& 0.02  & 0.05  & 0.02  & 0.10   & 0.02  & 0.07 \\
Background 
$\Dt$ shape    & 0.01 &$<0.01$& 0.04  &$<0.01$& 0.02   & 0.03  & 0.04 \\
Tag-side 
interference& $<0.01$ &$<0.01$&$<0.01$&$<0.01$& 0.01   &$<0.01$&$<0.01$\\
\hline      
Total          &$\SphikzSyst$ 
                      &$\SkpkmksSyst$
                              &$\SfzeroksSyst$
                                      &$\SetapksSyst$
                                              &$\SomegaksSyst$
                                                       &$\SkstarzgmSyst$
                                                               &$\SkspizSyst$\\
    \end{tabular}
  \end{ruledtabular}
\label{tab:ssyserr} 
\end{table*}
\begin{table*}
  \caption{Summary of the systematic errors on $\cala$.}
  \begin{ruledtabular}
    \begin{tabular}{lrrrrrr}
           & $\phi\kz$ 
                      &$\kp\km\ks$ 
                               &$\fzero\ks$ 
                                       &$\eta'\ks$
                                                &$\omega\ks$
                                                          &$\ks\piz$\\
\hline
Vertex 
reconstruction & 0.03   & 0.04   & 0.04   & 0.03   & 0.04   & 0.04 \\
Flavor tagging &$<0.01$ &$<0.01$ & 0.01   & 0.01   &$<0.01$ & 0.01 \\
Resolution 
function       & 0.02   & 0.02   & 0.02   & 0.01   & 0.04   &$<0.01$\\
Physics 
parameter      &$<0.01$ &$<0.01$ &$<0.01$ &$<0.01$ &$<0.01$ &$<0.01$\\
Possible 
fit bias       & 0.01   & 0.01   & 0.03   & 0.01 &$^{+0.01}
                                                   _{-0.03}$& 0.01 \\
Background 
fraction       & 0.04   & 0.01   & 0.06   & 0.02   & 0.14   & 0.02 \\
Background 
$\Dt$ shape    & 0.03   &$<0.01$ & 0.01   &$<0.01$ & 0.03   & 0.01 \\
Tag-side 
interference   & 0.06   & 0.06   & 0.03   & 0.03   & 0.03   & 0.05\\
\hline      
Total          &$\AphikzSyst$ 
                        &$\AkpkmksSyst$
                                &$\AfzeroksSyst$
                                          &$\AetapksSyst$
                                                   &$\AomegaksSyst$
                                                            &$\AkspizSyst$\\
    \end{tabular}
  \end{ruledtabular}
\label{tab:asyserr} 
\end{table*}

\clearpage
\newpage
\clearpage
\newpage
\clearpage
\newpage

\end{document}

%% file: defs.tex
\def\bz{{B^0}}
\def\bzb{{\overline{B}{}^0}}
\def\bp{{B^+}}
\def\bm{{B^-}}
\def\kl{K_L^0}
\def\dE{{\Delta E}}
\def\mb{{M_{\rm bc}}}
\def\Dt{\Delta t}
\def\Dz{\Delta z}
\def\fol{f_{\rm ol}}
\def\fsig{f_{\rm sig}}
\newcommand{\sinbb}{{\sin2\phi_1}}
 
\newcommand{\ra}{\rightarrow}
\newcommand{\myindent}{\hspace*{2cm}}  
\newcommand{\fCP}{f_{CP}}
\def\fcp{\fCP}
\newcommand{\ftag}{f_{\rm tag}}
\newcommand{\zCP}{z_{CP}}
\newcommand{\tCP}{t_{CP}}
\newcommand{\ttag}{t_{\rm tag}}
\newcommand{\cala}{{\cal A}}
\newcommand{\cals}{{\cal S}}
\newcommand{\dm}{\Delta m_d}
\newcommand{\dmd}{\dm}
\def\taubz{{\tau_\bz}}
\def\taubp{{\tau_\bp}}
\def\ks{{K_S^0}}
\newcommand{\btosqq}{b \to s\overline{q}q}
\newcommand{\btosss}{b \to s\overline{s}s}
\newcommand*{\dwl}{\ensuremath{{\Delta w_l}}}
\newcommand*{\fq}{\ensuremath{q}}
\def\kz{{K^0}}
\def\kp{{K^+}}
\def\km{{K^-}}
\def\fzero{{f_0(980)}}
\def\pip{{\pi^+}}
\def\pim{{\pi^-}}
\def\piz{{\pi^0}}
\def\kstarz{{K^{*0}}}
\def\kstarp{{K^{*+}}}
\def\kstarm{{K^{*-}}}
\def\kstarpm{{K^{*\pm}}}
\def\kl{{K_L^0}}
\def\bbar{{\overline{B}}}
\def\ufs{{\Upsilon(4S)}}
\def\nev{{N_{\rm ev}}}
\def\nsig{{N_{\rm sig}}}
\def\Nev{\nev}
\def\nsigmc{{N_{\rm sig}^{\rm MC}}}
\def\nbkg{{N_{\rm bkg}}}

\def\jpsi{{J/\psi}}
\def\dminus{{D^-}}
\def\dplus{{D^+}}
\def\dsm{{D^{*-}}}
\def\dpm{{D^{*+}}}
\def\rhop{{\rho^+}}
\def\rhom{{\rho^-}}
\def\rhoz{{\rho^0}}
\def\dzero{{D^0}}
\def\dzerob{{\overline{D}{}^0}}

\def\dzb{{\overline{D}{}^0}}
\newcommand{\dslnu}{D^{*-}\ell^+\nu}
\newcommand{\bzdslnu}{\bz \to \dslnu}
\newcommand*{\fflv}{\ensuremath{{f_\textrm{flv}}}}
\newcommand{\thetabdl}{\theta_{B,D^*\ell}}
\newcommand{\cosbdl}{\cos\thetabdl}

\def\sperp{{S_{\perp}}}
\def\lsig{{\cal L}_{\rm sig}}
\def\lbkg{{\cal L}_{\rm bkg}}
\def\rsigbkg{{\cal R}_{\rm s/b}}
\def\calf{{\cal F}}
\def\rkpi{{\cal R}_{K/\pi}}

\def\mgg{M_{\gamma\gamma}}
\def\ppizcms{p_\piz^{\rm cms}}

\def\pbstar{p_B^{\rm cms}}

\newcommand*{\eeff}{\ensuremath{\epsilon_\textrm{eff}}}
\def\egcms{{E_\gamma^{\rm cms}}}

\def\lnsig{{\cal L}_{N_{\rm sig}}}
\def\lzero{{\cal L}_0}

\def\acpraw{{A_{CP}^{\rm raw}}}

%% file: sss04_defs.tex
\def\efftot{{0.30\pm 0.01}}

\def\sinbbWA{+0.726}
\def\sinbbERR{0.037}
\def\sinbbWAResult{\sinbbWA\pm\sinbbERR}

\def\SphiksResPrv{-0.96\pm0.50^{+0.09}_{-0.11}}
\def\AphiksResPrv{-0.15\pm0.29\pm0.07}
\def\SetapksResPrv{+0.43\pm0.27\pm0.05}
\def\AetapksResPrv{-0.01\pm0.16\pm0.04}
\def\SkpkmksResPrv{-0.51\pm0.26\pm0.05}
\def\AkpkmksResPrv{-0.17\pm0.16\pm0.04}

\def\Nevjpsikspm{xxxx} 
\def\Pjpsikspm{0.xx} \def\Nsigjpsikspm{xxxx\pm xx}
 \def\Nevjpsikszz{xxx} 
\def\Pjpsikszz{0.xx} \def\Nsigjpsikszz{xxx\pm xx} 
  \def\Nevjpsikl{xxxx}     
  \def\Pjpsikl{0.xx}    \def\Nsigjpsikl{xxxx\pm xx}
    \def\Nevphiks{221}       
   \def\Pphiks{0.63}     \def\Nsigphiks{139 \pm 14}
    \def\Nevphikl{207}       
   \def\Pphikl{0.17}     \def\Nsigphikl{36 \pm 15}
\def\Nevkpkmks{718}     
  \def\Pkpkmks{0.56}    \def\Nsigkpkmks{399\pm 28}
\def\NevkspizH{298}      
   \def\PkspizH{0.55}     \def\NsigkspizH{168\pm 16}
\def\NevkspizL{499}      
   \def\PkspizL{0.17}     \def\NsigkspizL{83\pm 18}
\def\Nevkstarzgm{92} 
\def\Pkstarzgm{0.65}  \def\Nsigkstarzgm{57 \pm 9}
\def\Nevetapks{842}   
  \def\Petapks{0.61}    \def\Nsigetapks{512 \pm 27}
\def\Nevomegaks{56}  
 \def\Pomegaks{0.56}    \def\Nsigomegaks{31 \pm 7}
\def\Nevfzeroetcks{178}  
 \def\Pfzeroetcks{0.58}    \def\Nsigfzeroetcks{102 \pm 12}
 \def\Pfzeroks{0.53}    \def\Nsigfzeroks{94 \pm 14}
%
\def\SjpsikzVal{+0.666}   \def\SjpsikzStat{0.046}   \def\SjpsikzSyst{x.xx}
\def\AjpsikzVal{+0.023}   \def\AjpsikzStat{0.031}   \def\AjpsikzSyst{x.xx}
\def\SphikzVal{+0.06}    \def\SphikzStat{0.33}    \def\SphikzSyst{0.09}
\def\AphikzVal{+0.08}    \def\AphikzStat{0.22}    \def\AphikzSyst{0.09}
\def\SphiksVal{-x.xx}    \def\SphiksStat{x.xx}    \def\SphiksSyst{x.xx}
\def\AphiksVal{-x.xx}    \def\AphiksStat{x.xx}    \def\AphiksSyst{x.xx}
\def\SphiklVal{-x.xx}    \def\SphiklStat{x.xx}    \def\SphiklSyst{x.xx}
\def\AphiklVal{-x.xx}    \def\AphiklStat{x.xx}    \def\AphiklSyst{x.xx}
\def\SkpkmksVal{-0.49}   \def\SkpkmksStat{0.18}   \def\SkpkmksSyst{0.04}
\def\AkpkmksVal{-0.08}   \def\AkpkmksStat{0.12}   \def\AkpkmksSyst{0.07}
\def\SkpkmksFCP{^{+0.18}_{-0.00}}
\def\SkspizVal{+0.30}    \def\SkspizStat{0.59}    \def\SkspizSyst{0.11}
\def\AkspizVal{-0.12}    \def\AkspizStat{0.20}    \def\AkspizSyst{0.07}
\def\SkstarzgmVal{-0.79} \def\SkstarzgmStat{^{+0.63}_{-0.50}} \def\SkstarzgmSyst{0.10}
\def\AkstarzgmVal{-0.00} \def\AkstarzgmStat{0.38} \def\AkstarzgmSyst{x.xx}
\def\SetapksVal{+0.65}   \def\SetapksStat{0.18}   \def\SetapksSyst{0.04}
\def\AetapksVal{-0.19}   \def\AetapksStat{0.11}   \def\AetapksSyst{0.05}
\def\SomegaksVal{+0.75}  \def\SomegaksStat{0.64}  \def\SomegaksSyst{^{+0.13}_{-0.16}}
\def\AomegaksVal{+0.26}  \def\AomegaksStat{0.48}  \def\AomegaksSyst{0.15}
\def\SfzeroksVal{+0.47}  \def\SfzeroksStat{0.41}  \def\SfzeroksSyst{0.08}
\def\AfzeroksVal{-0.39}  \def\AfzeroksStat{0.27}  \def\AfzeroksSyst{0.08}
\def\SbsqqVal{+0.43} \def\SbsqqErr{^{+0.12}_{-0.11}}

\def\SjpsikzResult{\SjpsikzVal\pm\SjpsikzStat\pm\SjpsikzSyst}
\def\SjpsikzResultSS
  {\SjpsikzVal\pm\SjpsikzStat\mbox{(stat)}\pm\SjpsikzSyst\mbox{(syst)}}
\def\AjpsikzResult{\AjpsikzVal\pm\AjpsikzStat\pm\AjpsikzSyst}
\def\AjpsikzResultSS
  {\AjpsikzVal\pm\AjpsikzStat\mbox{(stat)}\pm\AjpsikzSyst\mbox{(syst)}}
\def\SphikzResult{\SphikzVal\pm\SphikzStat\pm\SphikzSyst}
\def\SphikzResultSS
  {\SphikzVal\pm\SphikzStat\mbox{(stat)}\pm\SphikzSyst\mbox{(syst)}}
\def\AphikzResult{\AphikzVal\pm\AphikzStat\pm\AphikzSyst}
\def\AphikzResultSS
  {\AphikzVal\pm\AphikzStat\mbox{(stat)}\pm\AphikzSyst\mbox{(syst)}}
\def\SphiksResult{\SphiksVal\pm\SphiksStat\pm\SphiksSyst}
\def\SphiksResultSS
  {\SphiksVal\pm\SphiksStat\mbox{(stat)}\pm\SphiksSyst\mbox{(syst)}}
\def\AphiksResult{\AphiksVal\pm\AphiksStat\pm\AphiksSyst}
\def\AphiksResultSS
  {\AphiksVal\pm\AphiksStat\mbox{(stat)}\pm\AphiksSyst\mbox{(syst)}}
\def\SphiklResult{\SphiklVal\pm\SphiklStat\pm\SphiklSyst}
\def\SphiklResultSS
  {\SphiklVal\pm\SphiklStat\mbox{(stat)}\pm\SphiklSyst\mbox{(syst)}}
\def\AphiklResult{\AphiklVal\pm\AphiklStat\pm\AphiklSyst}
\def\AphiklResultSS
  {\AphiklVal\pm\AphiklStat\mbox{(stat)}\pm\AphiklSyst\mbox{(syst)}}
\def\SkpkmksResult{\SkpkmksVal\pm\SkpkmksStat\pm\SkpkmksSyst}
\def\SkpkmksResultSS
  {\SkpkmksVal\pm\SkpkmksStat\mbox{(stat)}\pm\SkpkmksSyst\mbox{(syst)}}
\def\AkpkmksResult{\AkpkmksVal\pm\AkpkmksStat\pm\AkpkmksSyst}
\def\AkpkmksResultSS
  {\AkpkmksVal\pm\AkpkmksStat\mbox{(stat)}\pm\AkpkmksSyst\mbox{(syst)}}
\def\SkspizResult{\SkspizVal\pm\SkspizStat\pm\SkspizSyst}
\def\SkspizResultSS
  {\SkspizVal\pm\SkspizStat\mbox{(stat)}\pm\SkspizSyst\mbox{(syst)}}
\def\AkspizResult{\AkspizVal\pm\AkspizStat\pm\AkspizSyst}
\def\AkspizResultSS
  {\AkspizVal\pm\AkspizStat\mbox{(stat)}\pm\AkspizSyst\mbox{(syst)}}
\def\SkstarzgmResult{\SkstarzgmVal\SkstarzgmStat\pm\SkstarzgmSyst}
\def\SkstarzgmResultSS
  {\SkstarzgmVal\SkstarzgmStat\mbox{(stat)}\pm\SkstarzgmSyst\mbox{(syst)}}
\def\AkstarzgmResult{\AkstarzgmVal\pm\AkstarzgmStat\pm\AkstarzgmSyst}
\def\AkstarzgmResultSS
  {\AkstarzgmVal\pm\AkstarzgmStat\mbox{(stat)}\pm\AkstarzgmSyst\mbox{(syst)}}
\def\SetapksResult{\SetapksVal\pm\SetapksStat\pm\SetapksSyst}
\def\SetapksResultSS
  {\SetapksVal\pm\SetapksStat\mbox{(stat)}\pm\SetapksSyst\mbox{(syst)}}
\def\AetapksResult{\AetapksVal\pm\AetapksStat\pm\AetapksSyst}
\def\AetapksResultSS
  {\AetapksVal\pm\AetapksStat\mbox{(stat)}\pm\AetapksSyst\mbox{(syst)}}
\def\SomegaksResult{\SomegaksVal\pm\SomegaksStat\SomegaksSyst}
\def\SomegaksResultSS
  {\SomegaksVal\pm\SomegaksStat\mbox{(stat)}\SomegaksSyst\mbox{(syst)}}
\def\AomegaksResult{\AomegaksVal\pm\AomegaksStat\pm\AomegaksSyst}
\def\AomegaksResultSS
  {\AomegaksVal\pm\AomegaksStat\mbox{(stat)}\pm\AomegaksSyst\mbox{(syst)}}
\def\SfzeroksResult{\SfzeroksVal\pm\SfzeroksStat\pm\SfzeroksSyst}
\def\SfzeroksResultSS
  {\SfzeroksVal\pm\SfzeroksStat\mbox{(stat)}\pm\SfzeroksSyst\mbox{(syst)}}
\def\AfzeroksResult{\AfzeroksVal\pm\AfzeroksStat\pm\AfzeroksSyst}
\def\AfzeroksResultSS
  {\AfzeroksVal\pm\AfzeroksStat\mbox{(stat)}\pm\AfzeroksSyst\mbox{(syst)}}
\def\SbsqqResult{\SbsqqVal\SbsqqErr}

%% file: ichep04authors.tex
\affiliation{Aomori University, Aomori}
\affiliation{Budker Institute of Nuclear Physics, Novosibirsk}
\affiliation{Chiba University, Chiba}
\affiliation{Chonnam National University, Kwangju}
\affiliation{Chuo University, Tokyo}
\affiliation{University of Cincinnati, Cincinnati, Ohio 45221}
\affiliation{University of Frankfurt, Frankfurt}
\affiliation{Gyeongsang National University, Chinju}
\affiliation{University of Hawaii, Honolulu, Hawaii 96822}
\affiliation{High Energy Accelerator Research Organization (KEK), Tsukuba}
\affiliation{Hiroshima Institute of Technology, Hiroshima}
\affiliation{Institute of High Energy Physics, Chinese Academy of Sciences, Beijing}
\affiliation{Institute of High Energy Physics, Vienna}
\affiliation{Institute for Theoretical and Experimental Physics, Moscow}
\affiliation{J. Stefan Institute, Ljubljana}
\affiliation{Kanagawa University, Yokohama}
\affiliation{Korea University, Seoul}
\affiliation{Kyoto University, Kyoto}
\affiliation{Kyungpook National University, Taegu}
\affiliation{Swiss Federal Institute of Technology of Lausanne, EPFL, Lausanne}
\affiliation{University of Ljubljana, Ljubljana}
\affiliation{University of Maribor, Maribor}
\affiliation{University of Melbourne, Victoria}
\affiliation{Nagoya University, Nagoya}
\affiliation{Nara Women's University, Nara}
\affiliation{National Central University, Chung-li}
\affiliation{National Kaohsiung Normal University, Kaohsiung}
\affiliation{National United University, Miao Li}
\affiliation{Department of Physics, National Taiwan University, Taipei}
\affiliation{H. Niewodniczanski Institute of Nuclear Physics, Krakow}
\affiliation{Nihon Dental College, Niigata}
\affiliation{Niigata University, Niigata}
\affiliation{Osaka City University, Osaka}
\affiliation{Osaka University, Osaka}
\affiliation{Panjab University, Chandigarh}
\affiliation{Peking University, Beijing}
\affiliation{Princeton University, Princeton, New Jersey 08545}
\affiliation{RIKEN BNL Research Center, Upton, New York 11973}
\affiliation{Saga University, Saga}
\affiliation{University of Science and Technology of China, Hefei}
\affiliation{Seoul National University, Seoul}
\affiliation{Sungkyunkwan University, Suwon}
\affiliation{University of Sydney, Sydney NSW}
\affiliation{Tata Institute of Fundamental Research, Bombay}
\affiliation{Toho University, Funabashi}
\affiliation{Tohoku Gakuin University, Tagajo}
\affiliation{Tohoku University, Sendai}
\affiliation{Department of Physics, University of Tokyo, Tokyo}
\affiliation{Tokyo Institute of Technology, Tokyo}
\affiliation{Tokyo Metropolitan University, Tokyo}
\affiliation{Tokyo University of Agriculture and Technology, Tokyo}
\affiliation{Toyama National College of Maritime Technology, Toyama}
\affiliation{University of Tsukuba, Tsukuba}
\affiliation{Utkal University, Bhubaneswer}
\affiliation{Virginia Polytechnic Institute and State University, Blacksburg, Virginia 24061}
\affiliation{Yonsei University, Seoul}
  \author{K.~Abe}\affiliation{High Energy Accelerator Research Organization (KEK), Tsukuba} 
  \author{K.~Abe}\affiliation{Tohoku Gakuin University, Tagajo} 
  \author{N.~Abe}\affiliation{Tokyo Institute of Technology, Tokyo} 
  \author{I.~Adachi}\affiliation{High Energy Accelerator Research Organization (KEK), Tsukuba} 
  \author{H.~Aihara}\affiliation{Department of Physics, University of Tokyo, Tokyo} 
  \author{M.~Akatsu}\affiliation{Nagoya University, Nagoya} 
  \author{Y.~Asano}\affiliation{University of Tsukuba, Tsukuba} 
  \author{T.~Aso}\affiliation{Toyama National College of Maritime Technology, Toyama} 
  \author{V.~Aulchenko}\affiliation{Budker Institute of Nuclear Physics, Novosibirsk} 
  \author{T.~Aushev}\affiliation{Institute for Theoretical and Experimental Physics, Moscow} 
% ...
\collaboration{The Belle Collaboration}

%% file: ichep04ack.tex
\section*{Acknowledgments}
We thank the KEKB group for the excellent operation of the
accelerator, the KEK Cryogenics group for the efficient
operation of the solenoid, and the KEK computer group and
the National Institute of Informatics for valuable computing
and Super-SINET network support. We acknowledge support from
the Ministry of Education, Culture, Sports, Science, and
Technology of Japan and the Japan Society for the Promotion
of Science; the Australian Research Council and the
Australian Department of Education, Science and Training;
the National Science Foundation of China under contract
No.~10175071; the Department of Science and Technology of
India; the BK21 program of the Ministry of Education of
Korea and the CHEP SRC program of the Korea Science and
Engineering Foundation; the Polish State Committee for
Scientific Research under contract No.~2P03B 01324; the
Ministry of Science and Technology of the Russian
Federation; the Ministry of Education, Science and Sport of
the Republic of Slovenia; the National Science Council and
the Ministry of Education of Taiwan; and the U.S.\
Department of Energy.


%% file: sss04.bbl
\begin{thebibliography}{999}
\bibitem{bib:CC}
Throughout this paper,
the inclusion of the charge conjugate decay mode is implied
unless otherwise stated.
\bibitem{bib:BBbg}
Parameters in ${\cal P}_{\rm bkg}(\Dt)$ and $R_{\rm bkg}$
for $B\overline{B}$ background events in $\bz\to\eta'\ks$ decay
are determined from MC simulation.
\bibitem{footnote:phikz}
We use an alternative definition of
the raw asymmetry, 
$(N_{q\xi_f = -1}-N_{q\xi_f = +1})/(N_{q\xi_f = -1}+N_{q\xi_f = +1})$, 
for the decay $\bz\to\phi\kz$ to take the opposite $CP$ parities
for the decays $\bz\to\phi\ks$ and $\phi\kl$ into account.
\end{thebibliography}
